\documentclass[graphicx,aps,pre,epsfig,showpacs]{revtex4}
\usepackage{graphicx}
\usepackage{amsmath}

\begin{document}

\title{Polynomial iterative algorithms for coloring and analyzing random 
graphs}
\author{A. Braunstein}
\affiliation{International Center for Theoretical Physics, Strada
Costiera 11, P.O. Box 586, I-34100 Trieste, Italy}
\affiliation{SISSA, via Beirut 9, I-34100 Trieste, Italy}

\author{R. Mulet}
\affiliation{International Centre for Theoretical Physics, Strada
Costiera 11, P.O. Box 586, 34100 Trieste, Italy}
\affiliation{``Henri-Poincar\'e-Chair'' of Complex Systems and
Superconductivity Laboratory, Physics Faculty-IMRE, University of 
Havana, La Habana, CP 10400, Cuba}

\author{A. Pagnani}
\affiliation{International Centre for Theoretical Physics, Strada
Costiera 11, P.O. Box 586, 34100 Trieste, Italy}
\affiliation{Laboratoire de Physique Th\'eorique et Mod\`eles Statistiques,
b\^at. 100, Universit\'e Paris-Sud, F--91405 Orsay, France.}

\author{M. Weigt}
\affiliation{International Centre for Theoretical Physics, Strada
Costiera 11, P.O. Box 586, 34100 Trieste, Italy}
\affiliation{Institute for Theoretical Physics, University of
G\"ottingen, Bunsenstr. 9, 37073 G\"ottingen, Germany}

\author{R. Zecchina}
\affiliation{International Centre for Theoretical Physics, Strada
Costiera 11, P.O. Box 586, 34100 Trieste, Italy}
\date{\today}

\begin{abstract}
We study the graph coloring problem over random graphs of finite
average connectivity $c$. Given a number $q$ of available colors, we
find that graphs with low connectivity admit almost always a proper
coloring whereas graphs with high connectivity are uncolorable.
Depending on $q$, we find the precise value of the critical average
connectivity $c_q$.  Moreover, we show that below $c_q$ there exist a
clustering phase $c\in [c_d,c_q]$ in which ground states spontaneously
divide into an exponential number of clusters.  Furthermore, we
extended our considerations to the case of single instances showing
consistent results. This lead us to propose a new algorithm able to
color in polynomial time random graphs in the hard but colorable
region, i.e when $c\in [c_d,c_q]$.
\end{abstract}

\pacs{89.20.Ff, 75.10.Nr, 05.70.Fh, 02.70.-c}

\maketitle

\section{Introduction}
The Graph Coloring problem is a very basic problem in
combinatorics \cite{GaJo} and in statistical physics \cite{WU}. Given
a graph, or a lattice, and given a number $q$ of available colors, the
problem consists in finding a coloring of all vertices such that no edge
has the two ending vertices of the same color. The minimally needed
number of colors is the {\it chromatic number} of the graph.

For planar graphs there exists a famous theorem~\cite{four_colors}
showing that four colors are sufficient, and that a coloring can be
found by an efficient algorithm, while for general graphs the problem
is computationally hard to solve. In 1972 it was shown that Graph
Coloring is NP-complete~\cite{Karp72} which means, roughly speaking,
that the time required for determining the existence of a proper
coloring grows exponentially with the graph size. On the other hand if
an efficient algorithm for solving coloring in its worst case
instances exists, the same algorithm up {\em polynomial-reduction} can
be applied to efficiently solve all other problems in the class NP (for
a physicist's approach to complexity theory see~\cite{Mer}).

In modern computer science, graph coloring is taken as one of the most
widely used benchmarks for the evaluation of algorithm performance
\cite{satlib}. The interest in coloring stems from the fact that many
real-world combinatorial optimization problems have component
sub-problems which can be easily represented as coloring problems. For
instance, a classical application is the scheduling of registers in
the central processing unit of computers \cite{register}.  All
variables manipulated by the program are characterized by ranges of
times during which their values are left unchanged.  Any two variables
that change during the same time interval cannot be stored in the same
register.  One may represent the overall computation by constructing a
graph where each variable is associated with a vertex and edges are
placed between any two vertices whose corresponding variables change
during the same time interval.  A proper coloring with a minimal
number of colors of this graph provides an optimal scheduling for
registers: two variables with the same color will not be connected by
an edge and so can be assigned to the same register (since they change
in different time intervals).

The $q$-coloring problem of random graphs represents a very active
field of research in discrete mathematics which constitutes the
natural evolution of the percolation theory initiated by Erd\"os and
R\`enyi in the 50's \cite{Erdos_Renyi}.  One point of contact
between computer science and random graph theory arises from the
observation that, for large random graphs, there exists a critical
average connectivity beyond which the graphs become uncolorable by $q$
colors with
probability tending to one as the graph size goes to infinity. This
transition will be called the $q$-COL/UNCOL transition throughout this
paper. The precise value of the critical connectivity depends of
course on the number $q$ of allowed colors and on the ensemble of
random graphs under consideration.  Graphs generated close to their
critical connectivity are extraordinarily hard to color and therefore
the study of critical instances is at the same time a well posed
mathematical question as well as an algorithmic challenge for the
understanding of the onset of computational complexity \cite{AI,TCS}.
The notion of computational complexity refers to worst-case instances
and therefore results for a given ensemble of problems might not be of
direct relevance.  However, on the more practical side, algorithms
which are used to solve real-world problems display a huge variability
of running times and a theory for their typical-case behavior, on
classes of non-trivial random instances, constitutes the natural
complement to the worst-case analysis. Similarly to what happens for
other very famous combinatorial problem, e.g. the satisfiability
problem of Boolean formulas, critical random instances of $q$-coloring
are a popular test-bed for the performance of search
algorithms~\cite{satlib}.

From the physics side $q$-coloring has a direct interpretation as a
spin-glass model \cite{BOOKS_SG}. A proper coloring of a graph is
simply a zero-energy ground state configuration of a Potts
anti-ferromagnet with $q$-state variables. For most lattices such a
system is frustrated and displays all the equilibrium and
out-of-equilibrium features of spin glasses (the `Potts glass').

Here we focus on the $q$-coloring problem (or Potts anti-ferromagnet)
over random graphs of finite average connectivity, given by the ${\cal
G}(N,p)$ ensemble:  Graphs are composed of $N$ vertices, every pair of
them independently being connected by an edge with probability $p$,
and being not directly connected with probability $1-p$. The relevant
case of finite connectivity graphs is described by $p=c/N$, with $c$
staying constant in the thermodynamic limit $N\to\infty$. In this
case, the expected number of edges becomes $M=c/N {N\choose 2}\simeq
cN/2$,  i.e. proportional to the vertex number. Each of the vertices
is, on average, connected to $c$ other vertices. This connectivity
fluctuates according to a Poissonian, i.e. the probability of randomly
selecting a vertex with exactly $d$ neighbors is given by $e^{-c} c^d
/ d!$.

Two types of questions can be asked.  One type is algorithmic, i.e.
finding an algorithm that decides whether a given graph is
colorable. The other type is more theoretical and amounts to asking
whether a typical problem instance is colorable or not and what the
typical structure of the solution space is.  Here we address both
questions using the so called cavity method~\cite{MePa}. First we
provide a detailed description of the analytical calculations beyond
the results presented in~\cite{MuPaWeZe}, where the question of the
coloring threshold and of typical solution properties were
addressed. Second this analytical description is modified and applied
to single graph instances. This leads to an efficient graph coloring
algorithm for the region slightly below the COL/UNCOL transition. In
this region, known complete and stochastic algorithms are known to
fail already for moderate system sizes.

Let us start with reviewing some known results on the $q$-COL/UNCOL
transition on random graphs. One of the first important
finite-connectivity results was obtained by Luczak about one decade
ago \cite{Lu}. He proved that the threshold asymptotically grows like
$c_q \sim 2q\ln q$ for large numbers of colors, a result, which up to
a pre-factor coincides with the outcome of a replica calculation on
highly connected graphs \cite{KaSo} ($p={\cal O}(1)$ for large
$N$). For fixed number $q$ of colors, all vertices with less than $q$
neighbors, i.e. of {\it degree} smaller $q$, can be colored for
sure. The hardest to color structure is thus given by the maximal
subgraph having minimal degree at least $q$, the so-called
$q$-core. Pittel, Spencer and Wormald \cite{PiSpWo} showed that the
emergence of a 2-core coincides with the percolation transition of
random graphs at $c=1$ \cite{Erdos_Renyi}, and is continuous. For
$q\geq 3$, however, the $q$-core arises discontinuously, jumping from
zero to a finite fraction of the full graph. For $q=3$ they found
e.g. that the core emerges at $c\simeq 3.35$ and immediately contains
about 27\% of all vertices. Shortly after, it was realized that the
existence of the core is necessary, but in no way sufficient for
uncolorability \cite{Mo}. In fact, the best lower bound for the
3-COL/UNCOL transition is 4.03 \cite{AcMo1}, and numerical results
predict a threshold of about 4.7 \cite{Numerical}. The currently best
rigorous upper bound is 4.99 \cite{AcMo2}. It was obtained using a
refined first moment method. In statistical mechanics, the latter is
known as the annealed approximation. More recently, a replica
symmetric analysis of the problem has been performed~\cite{MoSa}. The
resulting threshold 5.1 exceeds, however, the rigorous bound, and one
has to go beyond replica symmetry. At the level of one-step
replica-symmetry breaking we are able to calculate a threshold value
$c_3 \simeq 4.69$ which we believe to be exact. We also describe the
solution space structure which undergoes a clustering transition at
$c_d \simeq 4.42$.

The remaining of the paper is organized as follows: In
section~\ref{sec:RS} we present the replica symmetric (RS) solution of
the problem and discuss why it fails. In section~\ref{sec:RSB} the
one-step replica-symmetry breaking (RSB) solution is presented. 
From this we derive the average graph connectivity for the q-COL/UNCOL 
transition, and we demonstrate the existence of a dynamic
threshold associated with the appearance of solution clusters in
configuration space. Then, in the next section we show that the
previous ideas are valid even in the analysis of single-case instances.
This allows, in section \ref{sec:ALG}, to propose an algorithm
that colors, in the hard region, single instances in polynomial time.
Finally, in Sec. \ref{sec:CONC} some conclusions of the work are presented.

\section{Replica Symmetric Solution}
\label{sec:RS}

As stated above, the question if a given graph is $q$-colorable
is equivalent to the question if there are zero-energy ground states
of the anti-ferromagnetic $q$-state Potts model defined on the same
graph. Denoting the set of all edges by $E$, the problem can thus
be described by the Hamiltonian
\begin{equation}
\label{eq_hamilton}
H_\mathcal{G} = \sum_{\{i,j\} \in E} \delta(\sigma_i,\sigma_j) 
\end{equation}
where $\{\sigma_i\} \in \{1,2,\dots,q\}$ are the usual Potts spins,
and $\delta(\cdot,\cdot)$ denotes the Kronecker symbol. This
Hamiltonian obviously counts the number of edges being colored equally
on both extremities, a proper coloring of the graph thus has energy
zero. Since this Hamiltonian cannot take negative values, the
combinatorial task of finding a coloring is translated to the physical
task of finding a zero-energy ground state, i.e. to the statistical
physics of the above model at zero temperature.

\subsection{The cavity equation}

In this paper we therefore apply the cavity method in a variant
recently developed for finite-connectivity graphs directly at zero
temperature \cite{MePaZe,MePa2,MeZe}. This approach consists of a
self-consistent iterative scheme which is believed to be exact over
locally tree-like graphs, like ${\cal G}(N,c/N)$, the set we consider
here. It includes the possibility of dealing with the existence of
many pure states. One has to first evaluate the energy shift of the
system due to the addition of a new new spin $\sigma_0$. Let us assume
for a moment that the new spin is only connected to a single spin, say
$\sigma_1$, in the pre-existing graph. Before adding the new site 0,
the ground-state energy of the system with fixed $\sigma_1$ can be
expressed as:
\begin{equation}
\label{eq:Es1}
E^N(\sigma_1) = A - \sum_{\tau=1}^q h^\tau_1 \delta(\tau,\sigma_1)
\end{equation}
where we have introduced the effective field $\vec
h_1=(h^1_1,...,h^q_1)$ and used the superscript $N$ to stress that the
previous quantity refers to the $N$-sites systems.  Note that a
$(q-1)$-dimensional field would be sufficient since one of the $q$
fields above can be absorbed in $A$. We, however, prefer to work with
$q$ field components in order to keep evident the global color
symmetry. When we connect $\sigma_0$ to $\sigma_1$ we can express the
minimal energy of the $N+1$-sites graph at fixed $\sigma_0, \sigma_1$
as:
\begin{equation}
\label{eq:Es0s1}
E^{N+1}(\sigma_0,\sigma_1) =  A - \sum_{\tau=1}^q h_1^\tau 
  \delta(\tau,\sigma_1) + \delta(\sigma_0,\sigma_1) 
\end{equation}
The minimum energy for the $N+1$-sites system at fixed $\sigma_0$ is
thus obtained by minimizing $E^{N+1}(\sigma_0,\sigma_1)$ with respect
to $\sigma_1$, it can be written as:
\begin{equation}
\label{eq:Es0}
E^{N+1}(\sigma_0) =\min_{\sigma_1}E^{N+1}(\sigma_0,\sigma_1) \equiv 
A - \omega (\vec h_1) - \sum_{\tau=1}^q \hat u^{\tau}(\vec h_1) \delta
(\tau,\sigma_0)
\end{equation}
with
\begin{eqnarray}
\label{eq:u}
\omega(\vec h) &=& -\min ( -h^1,..., -h^q )\\ \hat u^\tau(\vec h) &=&
-\min ( -h^1,...,-h^\tau+1,...,-h^q ) - \omega(\vec h)
\end{eqnarray}
where we have introduced the {\em cavity biases } $\hat u(\vec h)$.
This choice of $\omega$ and $\hat u$ is not unique but, according to
the previous discussion, we have chosen the only manifestly
color-symmetric notation. The structure of the cavity biases is easily
understood if we distinguish among two different cases:
\begin{itemize}
\item[(i)]  $ h^\tau  >  h^1 ,\dots,  h^{\tau-1},h^{\tau+1},\cdots,
  h^{q} $ for some $\tau$ : Then  $\hat u^\tau=  -1 $ and  $\hat u^\sigma=0$  for all
  $\sigma\neq \tau$.
\item[(ii)] $ h^{\tau_1} =  h^{\tau_2} \geq  h^1 ,\dots, h^q $ for
  some $\tau_1,\tau_2$ : 
  Then $\hat{u} = (\hat u^1,...,\hat u^q)=(0,0,\dots,0)=\vec 0$.
\end{itemize}
Only vectors $\vec h$ with non-degenerate maximal component give rise
to non trivial cavity bias $\vec u$ in the direction of the minimal
component. This is physically understandable: A unique maximal field
component in $\vec h$ fixes the corresponding color, which thus is
forbidden to the newly added site. If there are two or more maximal
field components, the color of the old site is not fixed, thus there
cannot be any forbidden color for the new vertex. Each cavity bias in
our problem thus belongs to the $q+1$-elements set $ \{ \vec{0}, \vec
e_1,\dots,\vec e_q \}$, where $\vec e_\tau$ has all components $0$ but
the $\tau^{th}$ equal to $-1$.

If the new spin $\sigma_0$ is connected to $k$ randomly chosen sites
with fields $\vec h_1, \dots, \vec h_k$, the cavity bias has to be
linearly superposed and the resulting cavity field on vertex 0 is
given by $\vec h_0 = \sum_{i=1}^k \hat u (\vec h_i ) $. With high
probability (tending to one for large $N$) the $k$ sites will be far
from each other in the original graph: Although an extensive number of
loops is surely present for $c>1$ \cite{Erdos_Renyi}, these loops have
lengths of the order $\log N$. Inside one Boltzmann state we can thus
invocate the {\em clustering propriety } \cite{BOOKS_SG} resulting in
a statistical independence of the $k$ selected sites and there cavity
fields $h_i$ (for a more detailed discussion of this point see
\cite{MeZe,MePa2}). The simplest ansatz assumes that there exists just
one such state (or a finite number, like in ferromagnets at low
temperature), which is equivalent to the Bethe-Peierls iterative
scheme or the replica-symmetric ansatz in the replica method.
Assuming further-on the existence of a well
defined thermodynamic limit of the energy density $E/N$ and of the
probability distributions of local fields (for recent rigorous studies
in this direction see \cite{GUERRA1,FraLeo,Tala}), the distribution of 
the fields $\vec h_0$ of the newly added vertex becomes the equal to
those of the $k$ neighbors. It is consequently determined by the
closed expression

\begin{figure}
\begin{center}
\includegraphics[angle=0,width=0.95\columnwidth]{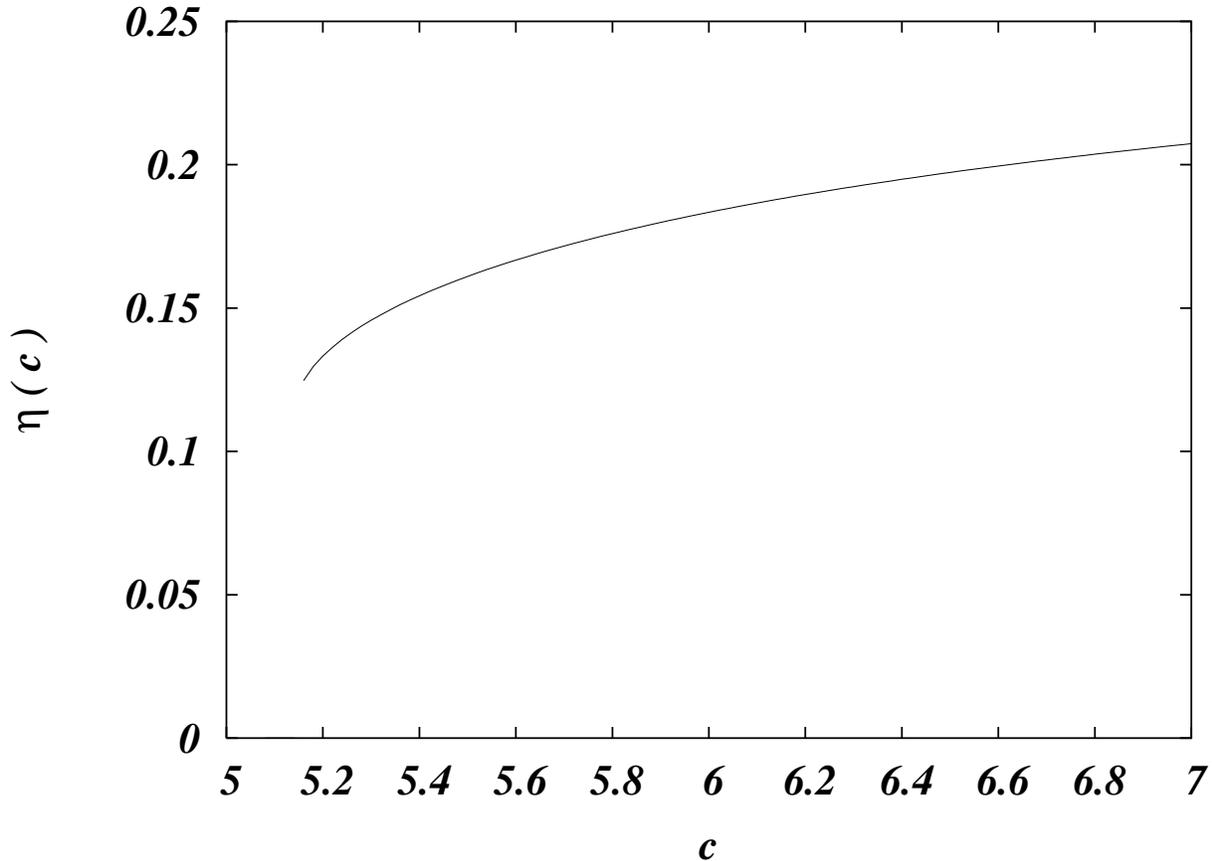}
\caption{Replica-symmetric order parameter $\eta$ vs. average
  connectivity $c$ for $q=3$, from Eq.~(\ref{eq_sprs})}
\label{fig_eta}
\end{center}
\end{figure}
\begin{eqnarray}
P(\vec{h}) &=& e^{-c} \sum_{k=0}^\infty \frac{c^k}{k!}
\int d^q \vec{h}_1, \dots ,d^q \vec{h}_n \,P(\vec{h}_1) \dots
P(\vec{h}_n)\, \delta ( \vec{h} - \sum_{i=0}^k \hat{u}_i(\vec{h}_i ) )
\\
\label{eq_QuPh}
Q(\vec{u}) &=&  \int d^q \vec h\,P(\vec{h})\,\delta ( \vec{u} -
\hat  u (\vec{h}))
\end{eqnarray}
where we have already used that the connectivities $k$ are distributed
according to a Poissonian of mean $c$.  The previous
equations can be combined in order to have a closed form for the
$Q(\vec{u})$:
\begin{equation}
\label{eq_Qu}
Q(\vec{u}) = e^{-c} \sum_{k=0}^\infty \frac{c^k}{k!} 
\int \prod_{i=1}^{k} d^q \vec{u_i}Q(\vec{u_i})
\delta(\vec{u} - \hat{u}(\vec{u_1} +\vec{u_2}  + \dots +\vec{u_k})).   
\end{equation}
From the symmetry of our model under arbitrary permutations of the colors
we conclude that
\begin{equation}
\label{eq_prob_u}
Q(\vec{e}_1)=Q(\vec{e}_2)=\dots=Q(\vec{e}_q)=\eta\,\,\,\,\,\,
        \mathrm{and}
\,\,\,\,\,Q(\vec{0}) = 1-q 
\end{equation}
{\em i.e.} we need a single real number $\eta$ with $0<\eta<\frac1q$
to completely specify the probability distribution function $Q(\vec
u)$. Noting that the probability $P^{(k)} (\vec h)$ for a site with $k$
neighbors can be expressed by
\begin{equation} P^{(k)}(\vec{h}) = \int \prod_{i=1}^k d^q \vec{u}_i
Q(\vec{u}_i) \delta (h-\sum_{i=1}^k \vec{u}_i)\ ,
\end{equation} 
and recalling that $\vec{u}_i \in
\{\vec{0},\vec{e}_1,\dots,\vec{e}_{q}\}$ it is easy to rewrite this
probability distribution in a compact multinomial form
\begin{equation}
\label{eq_Ph}
P^{(k)}(\vec{h}) \equiv P^{(k)} (h^1,h^2,\dots,h^{q}) = 
\frac { k! \,\,\,\eta ^{-\sum_{\tau=1}^{q} h^\tau} 
(1-q\eta )^{k+ \sum_{\tau=1}^{q} h^\tau}}
{ ( k + \sum_{\tau=1}^{q} h^\tau )! \prod_{\tau=1}^{q} (-h^\tau)! }
\end{equation}
with the convention that $1/n! = 0$ for $n < 0$. Note that $h^\tau \in
(0, -1, \dots, -k )$ and that there are correlations among the
different components of the cavity fields such that
$P^{(k)}(h^1,\dots,h^{q} ) \neq \prod_{\tau=1}^{q} {\cal P }(h^\tau)$.
Now we are ready to calculate the graph average over the Poissonian
connectivity distribution of mean $c$, 
\begin{equation}
P(h^1,...,h^{q})=  e^{-c}
\sum_{k} \frac{c^k}{ k!} P^{k}(h^1,\dots,h^{q}) =e^{-c\eta q}
\prod_{\tau=1}^{q} \frac{(c\eta)^{-h^\tau}}{(-h^\tau) !}  
\equiv \prod_{\tau=1}^q {\cal P}_{c\eta}(h^{\tau}) 
\label{eq_Phrs}
\end{equation}
It is interesting to note that after the average the correlations
among the different colors disappear and $P$ is the product of $q$
Poissonian distributions with average $c\eta$. From
Eq.~(\ref{eq_QuPh}) it is possible to derive a self-consistence
equation for the order parameter noting that the probability $\eta$ to
obtain a non-trivial cavity bias - say $\vec e_\tau$ - is simply given
by the probability that the $\tau^{\mathrm th}$ component of the local
field is the non-degenerate smaller, so setting $\tau=1$
\begin{equation}
\label{eq_sprs}
\eta = \sum_{h^1=0}^\infty \sum_{h^2=h1+1}^\infty \dots
\sum_{h^q=h1+1}^\infty P(h^1,\dots,h^q)= e^{-c\eta}
\sum_{n=0}^{\infty} \frac {(c\eta)^{n}}{n !}  \left( 1- \frac{\Gamma(n
+ 1, c\eta)}{\Gamma(n+1)} \right)^{q-1}
\end{equation}
where $\Gamma(n,x)$ is the incomplete Gamma function defined from the
following useful relation
\begin{equation}
 e^{-x}\,\sum_{k=n}^{\infty} \frac{x^k}{k!} = 1 - 
\frac{\Gamma(n,x)}{\Gamma(n)}
\end{equation}
The sum in Eq.~(\ref{eq_sprs}) converges very fast. It is therefore
easy to numerically construct a solution to this equation as a
function of $c$. For $q>2$ it turns out that $\eta$ jumps
discontinuously from zero to a finite value as shown in
Fig.~(\ref{fig_eta}) where the order parameter $\eta$ jumps at
$c=5.141$ in the case of $q=3$.
 
This means that, up to $c=5.141$ and at the level of the replica
symmetric assumption, we only find the paramagnetic solution $\eta =
0$. The solution $\eta > 0$ would account to a spontaneous breaking of
this symmetry, there should be a finite number of pure states (similar
to Neel order in Ising anti-ferromagnets).

\subsection{The calculation of the energy}

\begin{figure}
\begin{center}
\includegraphics[angle=0,width=0.95\columnwidth]{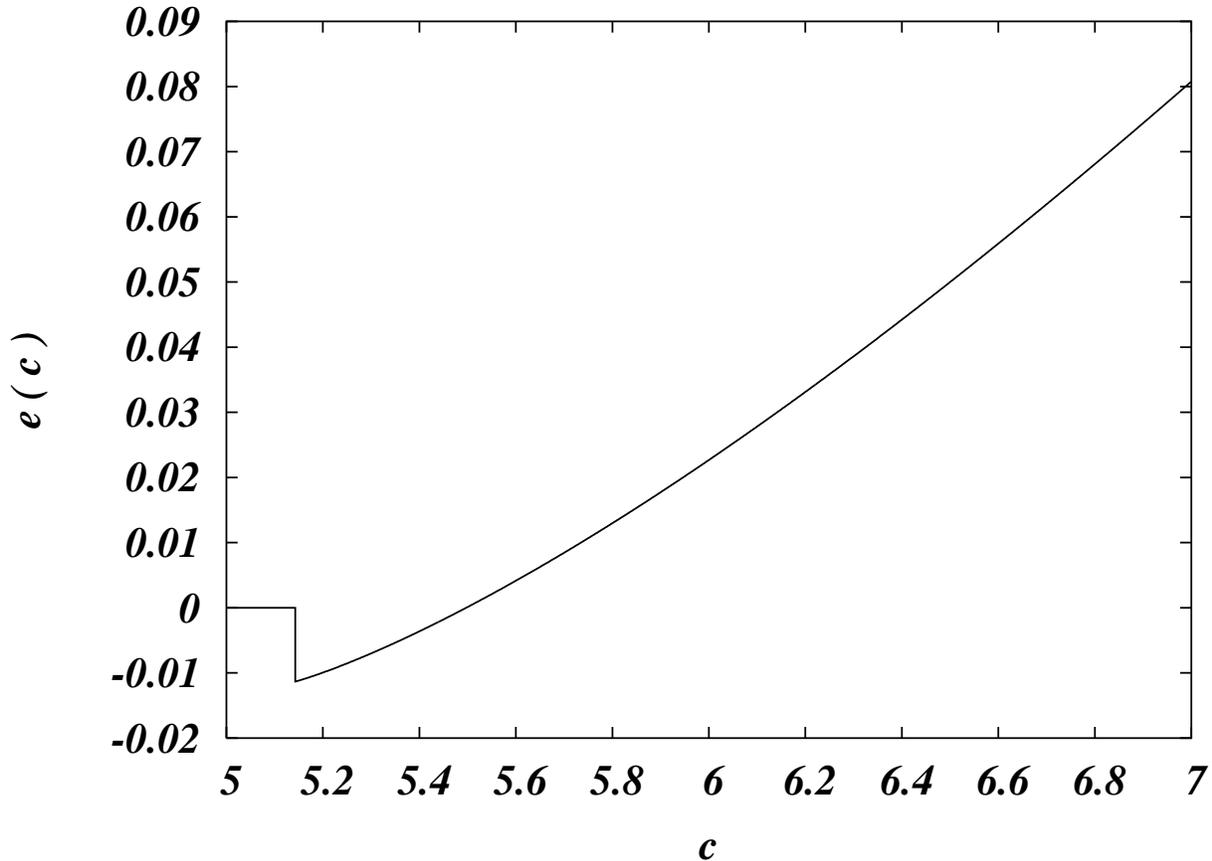}
\caption{Energy density $e$ vs. average connectivity $c$ for $q=3$ in
the RS approximation from Eq.~(\ref{eq_energia_rs})}
\label{fig_ene}
\end{center}
\end{figure}
One can easily compute the average shift in the ground state energy
when a new spin is added to the $N$-sites system and it is connected
to $k$ spins of the system. The energy of the original graph is given
by $A - \sum_{i=1}^k \omega(\vec h_i)$ while the energy of the
$N+1$-sites system is $A - \sum_{i=1}^k[ \omega(\vec h_i)+\omega(\hat
u (\vec h_i))]$. Therefore the average shift is given by  
\begin{equation}
\label{eq_de1}
\Delta E_1 = - \sum_{k=0}^\infty e^{-c} c^k/k!  \int d^q \vec u_1
\ldots d^q \vec u_k Q(\vec u_1)\dots dQ(\vec u_k) \,\omega\left(
\sum_{i=1}^k \vec u_i\right)= - \int d^q\vec h\, P(\vec h ) \,
\omega(\vec h)\ .
\end{equation}
One might be tempted to conclude that Eq.~(\ref{eq_de1}) equals the
energy density of the system, at least for $N$ large enough, but this
is not true. There is a correction term due to the change in the
number of links per variable in the iteration $N \rightarrow N+1$. In
fact, generating links with probability $c/N$ in a $N+1$ system,
instead of $c/(N+1)$ we are slightly over-generating links.  So, we
need to calculate the average energy shift in a system when two sites
- say spins $\sigma_1$ and $\sigma_2$ - are joined by an
anti-ferromagnetic link.

Again, the energy of the original graph is $A -
\omega(\vec h_1) - \omega(\vec h_2)$, while the energy after the two
spins are joined is given by $A - \min_{\sigma_1,\sigma_2}(-
h_1^{\sigma_1} - h_2^{\sigma_2} + \delta(\sigma_1,\sigma_2))$. The
difference between the two contributions can be written as 
\begin{eqnarray}
\label{eq_de2}
\Delta E_{\mathrm{link}} &=& 
\min_{\sigma_1} ( - h_1^{\sigma_1} + \min_{\sigma_2} ( - h_2^{\sigma_2} +
\delta(\sigma_1,\sigma_2) ) ) + \omega(\vec h_1 ) + \omega(\vec h_2 )
\nonumber \\
&=& \min_{\sigma_1}(- h_1^{\sigma_1} - u^{\sigma_1}(\vec h_2) -
\omega(\vec h_2 ) ) +   \omega(\vec h_1 ) + \omega(\vec h_2 )
\nonumber \\
&=& - \omega(\vec h_1 + \hat u(\vec h_2) ) + \omega(\vec h_1 )
\end{eqnarray}
This allows as to express the average link-energy shift as 
\begin{equation}
\label{eq_ene2}
\Delta E_2 = \int d^q \vec h_1 d^q \vec h_2 \, P(\vec h_1) P(\vec
h_2)\,\left( \omega(\vec h_1 ) - \omega(\vec h_1 + \hat u(\vec h_2) )
\right)
\end{equation}
It is interesting to observe that
Eqs.~(\ref{eq_de1})~and~(\ref{eq_de2}) are {\em model-independent}, in
the sense that the actual Hamiltonian is encoded into the functions
$\omega(\vec h)$ and $\hat u( \vec h )$ defined by Eq.~(\ref{eq:u}).

Using Eq.~(\ref{eq:u}) and (\ref{eq_Phrs}) one shows easily that 
Eq.~(\ref{eq_de1}) reduces to:
\begin{eqnarray}
\label{eq_ene1}
 \Delta E_1 &=&  \sum_{h^1 \dots h^{q}} {\cal P}_{c\eta}(h^1) \dots
        {\cal P}_{c\eta}(h^q) \min (- h^1,- h^2 ,\dots,- h^{q})
        \nonumber\\ &=& - \sum_{\alpha=0}^{q-1}\binom{q}{q-\alpha}
        \sum_{h=0}^{-\infty} h {\cal P }_{c\eta}(h)^{q-\alpha} \left(
        \sum_{g < h}^{-\infty} {\cal P}_{c\eta}(g) \right)^\alpha
\end{eqnarray}
It is also not hard to prove that the average link-energy shift $
\Delta E_2 = q \eta^2 $. This result can be obtained either by direct
computation of the integral, or following a simple probabilistic
argument: $\Delta E_{\mathrm{link}}$ is different from zero whenever
the two unlinked sites have the same color, but this happens with
probability $ \eta^2 $ for each of the colors.  Finally we have the
the following equation for the energy which is equivalent to the
replica-symmetric approximation:
\begin{equation}
\label{eq_energia_rs}
E = N \left (\Delta E_1 - \frac c2 \Delta E_2 \right) = -
        \sum_{\alpha=0}^{q-1}\binom{q}{q-\alpha} \sum_{h=0}^{-\infty} h
        {\cal P }_{c\eta}(h)^{q-\alpha} \left( \sum_{g < h}^{-\infty}
        {\cal P}_{c\eta}(g) \right)^\alpha \, - \,\frac c2  q \eta^2
\end{equation}
The behavior of the energy for $q=3$ as a function of the average
connectivity $c$ is displayed in Fig.~(\ref{fig_ene}). Let us note
that for average connectivity $5.141<c<5.497$ the energy is negative,
a particularly baffling result if we consider that the Hamiltonian
(\ref{eq_hamilton}) is at least positively defined. This phenomenon is
analogous to what already observed for the RS approximation in random
3-SAT \cite{MonZec}, and is a consequence of the approximation used.
We will see in the next section how the 1-RSB ansatz cures this
pathology. However, before leaving the RS approximation we would like
to compare our RS results with the RS approximation presented recently
by van Mourik and Saad in \cite{MoSa} since their result clearly
differs from ours. At the origin of the discrepancy is the fact that
they work a population dynamics at very low but finite temperature
finding a transition around $c=5.1$ but without negative energy
region. Analogously to what is reported in \cite{MePa2}, if one works
directly a zero temperature the distribution $P(\vec h)$ must be
concentrated around integer field components, but this is not true
anymore at temperature different from zero, as it happens in
\cite{MoSa}. They find that, in the zero-temperature limit, there
remain non-integer field components. In our opinion these are a direct
hint to the existence of RSB. Instead of including these fields into
an extended replica-symmetric approach, we directly switch to a
replica-symmetry broken solution.

\section{1 step RSB solution}
\label{sec:RSB}

The RS results show some evident pathologies and are at odd with
numerical simulations \cite{Numerical,MoSa} which predict a lower
threshold around $c=4.7$, and with the rigorous upper bound $c=4.99$
\cite{AcMo2}.  What can be wrong in our analysis? The main assumption
we have made is the statistical independence of the of the $k$ cavity
fields. Is it true that long distance among spins imply statistical
independence?  In general the answer we obtain from statistical
mechanics is no: The assumption holds only inside pure states.

In this section we will focus on how the cavity method could be used
to handle a situation in which there exist many different pure states.
More precisely we assume that their number ${\cal N}\propto
e^{N\Sigma}$ is exponentially large in $N$. The connectivity-dependent
exponent $\Sigma$ is called {\it complexity} and denotes the entropy
density of clusters. Note that it differs in general from the solution
entropy since each cluster may contain as well an exponential number of
solutions. The first basic assumption we made is that inside each pure
state the clustering condition holds. Under this assumption the
iteration can still be applied but we have to take into account the
reshuffling of energies of different states when new spins are added.

\subsection{ 1 RSB cavity equation }

We proceed following the same steps of the previous section. Let us
take the new spin $\sigma_0$ and let us connect it to $k$ spins
$\sigma_1,\dots, \sigma_k$ in the same state $\alpha$. Thanks to the
fast decrease of correlations inside a pure state the energy of state
$\alpha$ for fixed value of the $k$ spins is 
\begin{equation}
E_\alpha^{N}(\sigma_1,\dots,\sigma_k) = A_\alpha - \sum_{i=1}^k
\sum_{\tau=1}^q h^\tau_{i,\alpha} \delta(\tau,\sigma_{i,\alpha})
\end{equation} 
The optimization step within each pure state $\alpha$ runs still in
close analogy to the RS computation: when we connect $\sigma_0$ to
$\sigma_1,\dots,\sigma_k$, we express the minimal energy of the
$N+1$-sites graph with fixed $\sigma_0$, by minimizing the
$N+1$-sites system at fixed $\sigma_0$ is thus obtained by minimizing
$E^{N+1}_\alpha$ with respect to the $k$ spins:
\begin{equation}
E^{N+1}_\alpha(\sigma_0) = A_\alpha - \sum_{i=1}^k \omega(\vec
h_{i,\alpha} ) - \sum_{i=1}^k \sum_{\tau=1}^q \hat u^{\tau}(\vec 
h_{i,\alpha}) \delta (\tau,\sigma_0)
\end{equation}
This last equation shows that the local field acting on the new spin
$\sigma_0$ in the state $\alpha$ is 
\begin{equation}
\vec h_{0,\alpha} = \sum_{i=1}^k \hat u (\vec h_{i,\alpha} ) 
\end{equation}
and that the energy shift inside a state is 
\begin{equation}
\label{eq:shift}
\Delta E_\alpha = - \sum_{i=1}^k\omega\left(\hat u (\vec h_{i,\alpha}
  )\right)   
\end{equation}
All the previous equation are completely equivalent to those in the RS
case except for the fact that now we have a $\alpha$-index labeling
the different pure states. One natural question is how cavity fields
and the related cavity biases are distributed for a given site among
the different pure states. This leads us to the notion of {\em survey}
\cite{MePaZe, MePa2, MeZe}, {\em i.e.} the site dependent normalized
histogram over the different states of both cavity biases and cavity
fields:
\begin{eqnarray}
Q_i ( \vec u_i ) &=& \frac 1{\cal M }\sum_{\alpha=1}^{\cal M} 
\delta ( \vec u_i - \vec u_{i,\alpha}) \nonumber \\
P_i ( \vec h_i ) &=& \frac 1{\cal M }\sum_{\alpha=1}^{\cal M} 
\delta ( \vec h_i - \vec h_{i,\alpha})
\end{eqnarray}
In close analogy with what we have already done in the RS case, the
existence of a well defined thermodynamic limit implies that there
must exist unique functional probability distributions ${\cal Q}[
Q(\vec u)]$ and $ {\cal P}[ P(\vec u)]$ for all the surveys. One may
wander how could we handle such a big functional space: Fortunately
the $Q$-surveys are described in terms of a single real number
$0\leq\eta\leq 1/q$, cf. Eq.~(\ref{eq_prob_u}), and scalar function
$\rho(\eta)$ is enough for specifying their distribution:
\begin{equation}
\label{eq:dist-QQ}
{\cal Q}[ Q(\vec u)] = \int d \eta\ \rho(\eta)\  
\delta\left[ Q(\vec u) - (1-q\eta)\delta ( \vec u ) - \eta \sum_{\tau=1}^q
\delta (\vec u - \vec e_{\tau}) \right]
\end{equation}
with $\delta[\cdot]$ denoting a functional Dirac distribution.
Assuming that the survey of site $0$ is distributed equally to those
of all its k neighbors, we can write:
\begin{eqnarray}
\label{eq_surv_P}
P_0(\vec h) &=&  e^{-c} \sum_{k=0}^\infty \frac{c^k}{k!} C_k \int d^q
\vec u_1 Q_1(\vec u_1 ) \cdots  d^q \vec u_k Q_k(\vec u_k ) e^{ y
\omega ( \sum_{i=1}^k \vec u_i )} \delta ( \vec h - \sum_{i=1}^k \vec u_i )  
\\
\label{eq_surv_Q}
Q_0(\vec u) &=& \int d^q \vec h P_0(\vec h ) \delta ( \vec  u - \hat u
( \vec h ) )  
\end{eqnarray}
Note the presence of the {\em reweighting factor} $\exp ( y \omega (
\sum_{i=1}^k \vec u_i )) $ that arise after conditioning the
probability distributions of the $\vec{h}$s to a given value of
energy \cite{MePa2}, the prefactors $C_k$ are normalization constants
depending on $Q_1(\vec u),...,Q_k(\vec u)$. The reweighting parameter
$y$ is a number equal to
the derivative of the complexity $\Sigma(e)$ of metastable states 
with respect to their energy density $e=E/N$:
\begin{equation}
y = \frac{\partial \Sigma}{\partial e}
\end{equation} 
Intuitively, this reweighting factor can be understood as a penalty
$e^{-y\Delta E_\alpha}$ one has to pay for positive energy shifts. 
Note that Eqs.~(\ref{eq_surv_P})~and~(\ref{eq_surv_Q}) can be cast
in the following form
\begin{eqnarray}
\label{eq_sp_1rsb}
Q_0(\vec u_0) &=&  e^{-c} \sum_{k=0}^\infty \frac{c^k}{k!} C_k  
\int d^q \vec u_1 Q_1(\vec u_1) \cdots d^q \vec u_k Q_k(\vec u_k)
e^{y \omega (\sum_{i=1}^k \vec u_i)} \delta(\vec u_0 - \hat u (
\vec u_1+ \dots + \vec u_k)) \nonumber \\
&=& e^{-c} \sum_{k=0}^\infty \frac{c^k}{k!} C_k  
\int d^q \vec h \tilde P(\vec h ) e^{y \omega ( \vec h ) } 
\delta ( \vec u_0 - \hat u ( \vec h ))
\end{eqnarray}
In the last line we have introduced the auxiliary distribution $\tilde
P(\vec h)$ which would result in Eq. (\ref{eq_surv_P}) without
reweighting (i.e. by setting $y=0$). It has no direct physical meaning
in this context, but it will be of great technical help in the
following calculations.

\begin{figure}
\begin{center}
\includegraphics[angle=0,width=.94\columnwidth]{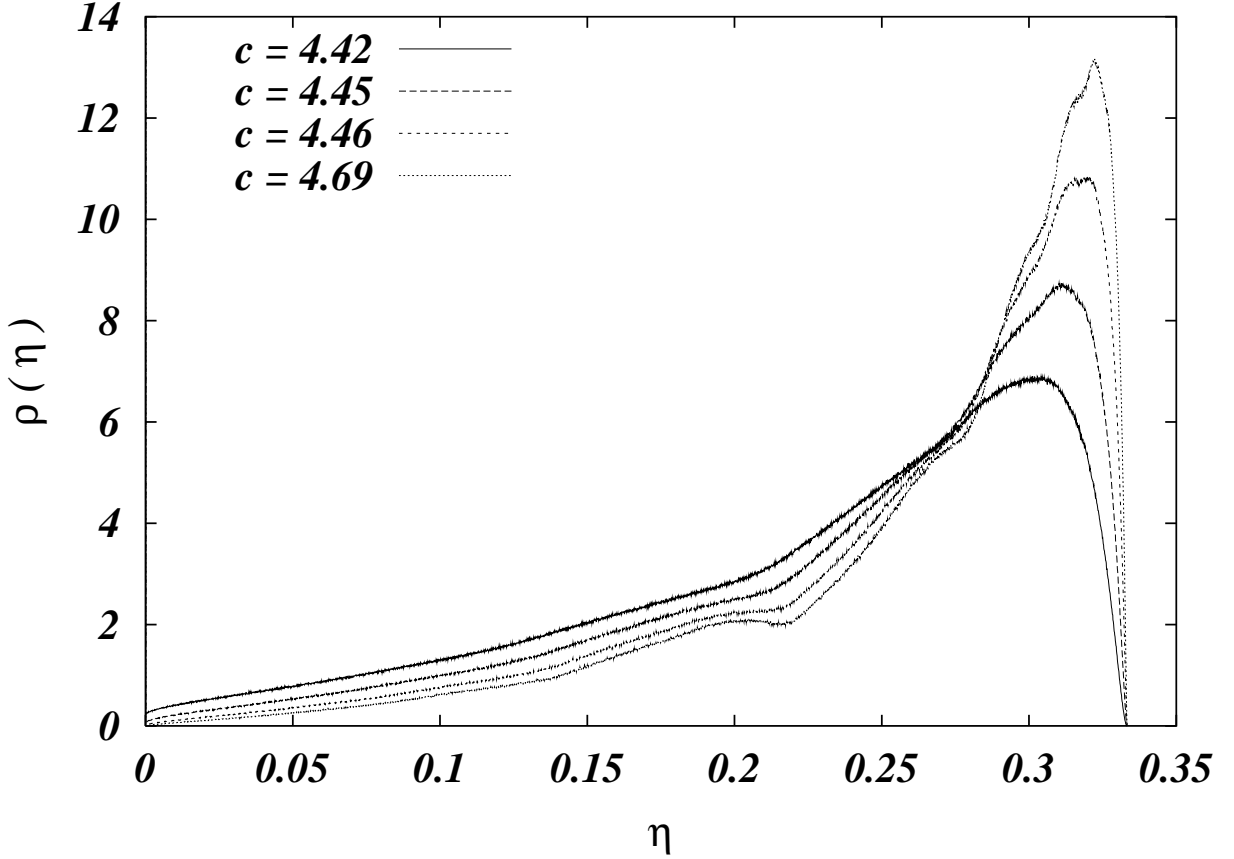}
\caption{Probability distribution function $\rho(\eta)$ in the case
$q=3$ for average connectivities $4.42<c<4.69$. 
Note also that a delta peak in $\eta=0$ is always present (but
not displayed here).}
\label{fig_rho_eta}
\end{center}
\end{figure}

Let us first concentrate on the {\it colorable phase}, where the
ground states are proper q-colorings and have zero energy. Consequently
no positive energy shifts are allowed, so this phase is characterized
by $y=\infty$. Let us first calculate the value of the normalization
constants $C_k$ in this limit. Note that $\omega(\vec h ) \leq 0 $ for
all allowed $\vec h$ (each component of $\vec h$ is non-positive as
$\vec h$ results from a sum over $\vec u$s). This means that the only
surviving terms in Eq.~(\ref{eq_sp_1rsb}) are those with zero energy
shift $\omega(\vec h )=0$, {\em i.e.} all fields must have at least
one zero component, allowing for the selecting of at least one color
without violating an edge. Let us first specialize to the case $q=3$
for clarity, the generalization to arbitrary $q$ is straightforward.
Summing over $\vec u_0$ both sides of Eq.~(\ref{eq_sp_1rsb}) we have:
\begin{equation}
\label{eq_C}
\frac 1{C_k} = \tilde P(0,0,0) + 3 \sum_{h^1 < 0} \tilde P(h^1,0,0) + 
3 \sum_{h^1,h^2 < 0} \tilde P(h^1,h^2,0)
\end{equation}
where the combinatorial factors 3 appearing in the r.h.s. are obtained
by noting that $\tilde P(h,0,0)=\tilde P(0,h,0)=\tilde P(0,0,h)$ and
that $\tilde P(h^1,h^2,0)=\tilde P(h^1,0,h^2)=\tilde P(0,h^1,h^2)$.
Combining Eqs.~(\ref{eq_surv_P}),(\ref{eq_surv_Q}) and (\ref{eq_prob_u})
we get
\begin{eqnarray}
\tilde P(0,0,0) &=& \prod_{i=1}^k (1-3\eta_i) 
\\
\sum_{h^1<0} \tilde P(h^1, 0, 0 ) &=& \prod_{i=1}^k ( 1 - 2 \eta_i ) -
\tilde P(0,0,0) =  \prod_{i=1}^k ( 1 - 2 \eta_i ) - 
\prod_{i=1}^k (1-3\eta_i)
\\
\label{eq_sp1rsb}
\sum _{h^1,h^2 < 0} \tilde P(h^1,h^2,0) &=& \prod_{i=1}^k (1 - \eta_i) 
- 2 \sum_{h^1<0} \tilde P(h^1, 0, 0 ) - \tilde P(0,0,0)  \nonumber\\ 
&=&  \prod_{i=1}^k (1-\eta_i) - 2  
\prod_{i=1}^k (1-2\eta_i) + \prod_{i=1}^k (1 - 3\eta_i)
\end{eqnarray} 
Plugging these relations into Eq.~(\ref{eq_C}) we finally get 
\begin{equation}
\label{eq_C_final}
\frac 1{C_k} = 3 \prod_{i=1}^k (1 - \eta_i)  - 3 \prod_{i=1}^k (1 - 2
\eta_i) +  \prod_{i=1}^k (1 - 3 \eta_i)  
\end{equation}
Note also that in close analogy to the analysis that lead to
Eq.~(\ref{eq_sprs}), we can interpret (\ref{eq_sp1rsb}) as the
(un-normalized) probability of having the survey in site $0$ {\em
pointing} in direction $\vec e_3$. Therefore combining
Eqs.~(\ref{eq_sp1rsb}) and (\ref{eq_C_final}) we obtain 
\begin{equation}
\label{eq_self_cons}
\eta_0 = \hat f_k ( \eta_1, \dots, \eta_k) = \frac
{\prod_{i=1}^k (1-\eta_i) - 2  
\prod_{i=1}^k (1-2\eta_i) + \prod_{i=1}^k (1 - 3\eta_i)}
{3 \prod_{i=1}^k (1 - \eta_i)  - 3 \prod_{i=1}^k (1 - 2
\eta_i) +  \prod_{i=1}^k (1 - 3 \eta_i)}
\end{equation}
At this point we are ready to write the 1-RSB iterative equation for
the $Q$-surveys:
\begin{equation}
\label{eq_sc1rsb}
\rho(\eta) = e^{-c} \sum_{k=0}^{\infty} \frac{c^k}{k!} 
\int d\eta_1\rho(\eta_1) \cdots d\eta_k\rho(\eta_k)\ 
\delta (\eta - \hat f_k(\eta_1, \dots,\eta_k))
\end{equation}
Eq.~(\ref{eq_self_cons}) can be easily generalized to an arbitrary
number $q$ of colors,
\begin{equation}
\label{eq_self_cons_q}
\hat f_k(\eta_1,...,\eta_k) = 
\frac{\sum_{l=0}^{q-1} (-1)^l {q-1 \choose l} \prod_{i=1}^k
[1-(l+1)\eta_i] }
{\sum_{l=0}^{q-1} (-1)^l {q \choose l+1} \prod_{i=1}^k
[1-(l+1)\eta_i] }
\end{equation}
The self-consistent equation (\ref{eq_sc1rsb}) resembles a
replica-symmetric equation and can be solved numerically using a
population dynamic algorithm: 
\begin{itemize}
\item[(i)] Start with an initial population $\eta_1,...,\eta_{\cal M}$ of
size ${\cal M}$ which can be easily chosen to be as large as $10^6$ to
generate high-precision data. 
\item[(ii)] Randomly draw a number $k$ from the Poisson distribution $e^{-c}
c^k/k!$; 
\item[(iii)] Randomly select $k+1$ indices $i_0,i_1,...,i_k$ from
$\{1,...,{\cal M}\}$; 
\item[(iv)] Update the population by replacing
$\eta_{i_0}$ by $f_d(\eta_{i_1},...,\eta_{i_k})$;
\item[(v)] {\tt GOTO} (ii) until convergence of the algorithm is reached.
\end{itemize}
One obvious solution of Eq.~(\ref{eq_sc1rsb}) and
~(\ref{eq_self_cons_q}) is the paramagnetic solution
$\delta(\eta)$. For small average connectivities $c$ it is even the
only one. The appearance of a non-trivial solution coincides with a
clustering transition of ground states into an exponentially large
number of extensively separated clusters. In spin-glass theory, this
transition is called dynamical.  Still, $\rho(\eta)$ will contain a
non-trivial peak in $\eta=0$ due to small disconnected subgraphs,
dangling ends, low-connectivity vertices etc. The shape of $\rho (
\eta )$ in the case $q=3$ is displayed in Fig.~(\ref{fig_rho_eta}) for
connectivities $c$ ranging from $c_d$ to $c_c$.

The weight $t$ of this peak can be computed self-consistently. Let us
again consider first the case $q=3$. Keeping in mind that for
$y\rightarrow\infty$ the field $\vec h$ has at least one vanishing
component, the only possibilities to obtain $\hat u ( \vec h ) = \vec
0$ are given by $\vec h= \vec 0$ or by a field $\vec h$ with one
single non-zero component. So the probability that the
cavity field acting on a given site with $k$ neighbors equals zero 
is given by the sum of the probabilities that all neighboring cavity 
fields are zero (equal to $t^k$), plus the probability that exactly 
one cavity bias among the $k$ is non-trivial (equal to $k
(1-t) t^{k-1}$). The average over the Poissonian degree distribution
leads to
\begin{equation}
\label{eq:3core}
t = e ^{-c}\sum_{k=0}^\infty \frac {c^k}{k!} 
    \big( t^k + k  t^{k-1} (1-t)\big)
  = e^{-(1-t)c} \big( 1 + (1-t) c \big)   
\end{equation}
Generalizing Eq.~(\ref{eq:3core}) to a general number $q$ of colors
easily gives 
\begin{equation}
\label{eq:qcore}
t= e^{-(1-t)c} \sum_{l=0}^{q-2} \frac{(1-t)^lc^l}{l!}\ .
\end{equation}
This equation is quite interesting, since a non-trivial solution forms
a necessary condition for Eq. (\ref{eq_sc1rsb}) to have a non-trivial
solution. In fact, this equation was first found in \cite{PiSpWo}, the
fraction of edges belonging to the $q$-core is given by $(1-t_{min})$
with $t_{min}$ being the smallest positive solution of Eq.
(\ref{eq:qcore}). Thus, we also find that the existence of an
extensive $q$-core is necessary for a non-trivial $\rho(\eta)$, and
forms a lower bound for the $q$-COL/UNCOL transition.

Unlike in the case of finite-connectivity $p$-spin-glasses or,
equivalently, random XOR-SAT problems \cite{RiWeZe,CoDuMo,MeRiZe}, the
existence of a solution $t<1$ is not sufficient for a non-trivial
$\rho(\eta)$ to exist. The latter appears suddenly at the dynamical
transition $c_d$, which can be determined to high precision using the
population dynamical algorithm. This solution does not imply
uncolorability, but the set of solutions is separated into an
exponentially large number of clusters. The number of these clusters,
or more precisely its logarithm divided by the graph size $N$, is
called the complexity $\Sigma$ and can be calculated from
$\rho(\eta)$.

\subsection{The calculation of energy and complexity}

More generally, we expect also a large number of metastable states at
non-zero energy to exist. Hereafter we will assume that they are
exponentially many, $\mathcal{N}(e) \propto \exp(N \Sigma(e))$, 
where the complexity $\Sigma(e)$ is (despite of the use of a capitol
letter!) an intensive function of the energy density $e=E/N$. We can
introduce a thermodynamic potential $\phi(y)$ \cite{Educated_Reader}
as

\begin{equation}
\label{eq_part_fun}
\phi(y) = -\frac 1{yN} \ln\Big(\int de\ e^{N\{-ye +
\Sigma(e)\} }\Big)
\end{equation}
For large $N$, we calculate this integral by its saddle point: 
\begin{equation}
\label{eq_legendre}
\phi(y) = \min_e \Big( e - \frac 1y \Sigma(e)\Big) = e_{sp} - \frac 1y
\Sigma(e_{sp})   
\end{equation}  
It is easily verified that the potential $\phi$ calculated at the
saddle point energy $e_{sp}(y)$ fulfills the usual Legendre
relations:
\begin{eqnarray}
\label{eq_esp}
\partial_y\left[y\,\phi(y,e_{sp}(y))\right] &=& e_{sp}\\        
\label{eq_complexity}
y^2 \partial_y \phi(y,e_{sp}(y)) &=& \Sigma(e_{sp})
\end{eqnarray}
Around the saddle point the complexity can be approximated, according
to equation (\ref{eq_legendre}), by
\begin{equation}
\Sigma(e) \simeq \Sigma(e_{sp}) + y(e - e_{sp}) = - y\phi(y) + ye
\end{equation} 
We will now consider a cavity argument: let us denote by $E_N$ the
energy of a system composed of $N$ sites, the density of
configurations is given by
\begin{equation} 
\label{eq_den_state}
d{\mathcal N}_{N}(E_N)\propto e ^{-y \Phi_N(y) + yE_N } dE_N
\end{equation} 
with $\Phi_N(y)$ denoting the extensive thermodynamic potential with
limit $\Phi_N(y)/N \to \phi(y)$. Now we add a spin to the system.  If
we consider that the total energy is $E_{N+1} = E_N + \Delta E$, we
can express the density of configurations in terms of $E_{N}$ and
$\Delta E$:
\begin{equation} 
\label{eq_en_add}
d{\mathcal N}_{N+1}(E_N,\Delta E) \propto e^{ y
\Phi_{N}(y)+y(E_N + \Delta E)}\ dE_N\ P(\Delta
E)\ d\Delta E\ .
\end{equation} 
Integrating over $\delta E$ we get
\begin{eqnarray} 
d{\mathcal N}_{N+1}(E_{N+1}) &=&  C e^{
-y \Phi_{N+1}(y)+yE_{N+1}}dE_{N+1}\\ 
C &=& \frac 1y \int
P(\Delta E)e^{y\Delta E} d\Delta E \equiv \frac1y \langle e^
{ y\Delta E}\rangle_{P(\Delta E)} 
\end{eqnarray}
Comparing the previous equations with (\ref{eq_den_state}) we
can deduce that
\begin{equation}
\Phi_{N+1}(y) = \Phi_N(y) - \frac 1y \ln
\langle \exp{(y\Delta E)}\rangle_{P(\Delta E)}\,\,. 
\end{equation} 
In the thermodynamic limit we can thus identify
\begin{equation}
\phi(y) = -  \frac 1y \ln \langle \exp{( y\Delta E)}
\rangle_{P(\Delta E)} 
\end{equation}
In close analogy with what we have already done in the RS case, and
using Eq. (\ref{eq:shift}), we can compute $\phi$ as a {\em site}
contribution plus a {\em link} contribution in the 1-RSB scenario:
\begin{itemize} 
\item{{\em Site Addition}}
\begin{equation}
\label{eq_de1_1rsb}
\exp \big(- y \Delta \phi_1 \big) = \int d^q \vec u_{i_1} 
Q_{i_1}(\vec u_{i_1}) \cdots d^q \vec u_{i_k} Q_{i_k}(\vec u_{i_k}) 
\exp\left( y \omega ( \sum_{j=1}^k \vec u_{i_j} ) \right) 
= \frac 1{C_k}   
\end{equation}

\item{{\em Link Addition}}
\begin{eqnarray}
\label{eq_de2_1rsb}
\exp \big(- y \Delta \phi_2 \big) &=& \int d^q \vec h_{i_1}
P_{i_1}(\vec h_{i_1}) \ d^q \vec h_{i_k} P_{i_2}(\vec h_{i_2}) 
\exp\Big( -y\omega(\vec h_{i_1} ) + y\omega\big(\vec h_{i_1 }+ 
\hat u(\vec h_{i_2})\big) \Big) 
\nonumber\\
&=& \int d^q\vec h P_{i_1}(\vec h) d^q  \vec u
Q_{i_2}(\vec u ) \exp [ - y \big ( \omega(\vec h) -  \omega (
\vec h + \vec u) \big)]  \nonumber\\
&=& 1 + q \eta_{i_1}\eta_{i_2}(e^{-y} - 1) 
\end{eqnarray}
\end{itemize}
Note that in the limit $y\rightarrow 0$ and assuming $P_i = P $ for
each site, we obtain the RS expressions. Once the functional
distributions ${\cal Q}[Q(\vec u )]$ and ${\cal P}[P(\vec h )]$ are
known we can eventually average the energy shifts $\Delta \phi_1$, 
$\Delta\phi_2$ in the usual linear combination:
\begin{equation}
\phi(y) = \overline{ \Delta\phi_1 } - \frac c2\overline{\Delta \phi_2}  
\end{equation}
where the over-lines  denote the average over both disorder and
functional distributions. One finally finds
\begin{eqnarray}
\phi(y) &=& -\frac 1y \sum_{k=1}^\infty e^{-c}\frac {c^k}{k!} 
\int \prod_{i=0}^k {\cal D}Q_i {\cal Q}[Q_i]\ \ \ln \left(\int
  \prod_{i=0}^k d^q \vec  u_i Q_i(\vec u_i)\  
\exp\left( y \omega ( \sum_{i=1}^k \vec u_{i} ) \right)  
 \right) +\nonumber\\ 
&& +\frac c{2y} \int {\cal D}P_1{\cal P}[P_1]\  {\cal D}P_2{\cal
  P}[P_2]   
\ln \left( \int  d^q\vec h_1 P_{1}(\vec h_{1}) \ d^q\vec h_2
P_{2}(\vec h_{2}) \exp\Big( y\omega(\vec h_{1} ) - 
y\omega\big(\vec h_{1}+
\hat u(\vec h_{2})\big) \Big)
\right) 
\end{eqnarray}
In the limit $y \rightarrow \infty$ these relations can be written in
a more explicit form. Let us consider first the term $\Delta \phi_1$
in Eq.~(\ref{eq_de1_1rsb}). Referring to Eq.~(\ref{eq_C_final}) it 
easy to see that:
\begin{equation}
\lim_{y\rightarrow\infty} e^{-y \Delta \phi_1 } =
\sum_{l=0}^{q-1} (-1)^l {q \choose l+1} \prod_{i=1}^k [1-(l+1)\eta_i]
\end{equation}
such that 
\begin{equation}
\lim_{y\rightarrow\infty} -y \overline{ \Delta \phi_1} = \sum_{k=1}^\infty
e^{-c}\frac {c^k}{k!}  \int d\rho(\eta_1) \dots d\rho(\eta_k)
\ln\left(\sum_{l=0}^{q-1} (-1)^l {q \choose l+1}
\prod_{i=1}^k[1-(l+1)\eta_i] \right). 
\end{equation}
In order to compute the average link contribution $\overline{\Delta
\phi_2(y)}$ we need to evaluate the large $y$ limit of
Eq.~(\ref{eq_de2_1rsb}) which gives:
\begin{equation}
\lim_{y\rightarrow\infty} -y \overline{\Delta \Phi_2(y)}
= \int d\rho(\eta_1) d\rho(\eta_2) \ln\left(1-q\eta_1\eta_2 \right)
\end{equation}  
This equation has a nice probabilistic interpretation complementary to
that used in the derivation of $\Delta E_2$ in the RS case. In fact
the integrand of (\ref{eq_de2_1rsb}) is different from zero for
$y\rightarrow \infty$ only when both sites $i_1$ and $i_2$ have a
different color, and this happens with probability
$(1-q\eta_{i_1}\eta_{i_2})$ (note that $q\eta_{i_1}\eta_{i_2}$ is the
probability that the two sites have same color).  It is now clear from
Eq.~(\ref{eq_legendre}) that taking the $y\rightarrow\infty$ of
$-y\Phi(y)$ gives us the complexity at least in the COL region where
$e=0$:
\begin{eqnarray}
\label{eq:sigma}
\Sigma(e=0) &=& \sum_{k=1}^\infty
e^{-c}\frac {c^k}{k!}  \int d\eta_1 \rho(\eta_1) \dots d\eta_k \rho(\eta_k)
\ln\left(\sum_{l=0}^{q-1} (-1)^l {q \choose l+1}
\prod_{i=1}^k[1-(l+1)\eta_i] \right) \nonumber\\
&& - \frac c2 \int d\eta_1 \rho(\eta_1) d\eta_2\rho(\eta_2) 
\ln\left(1-q\eta_1\eta_2 \right)
\end{eqnarray}

\subsection{Results}
\label{sub_res}
\begin{figure}[htb]
\includegraphics[width=0.94 \columnwidth]{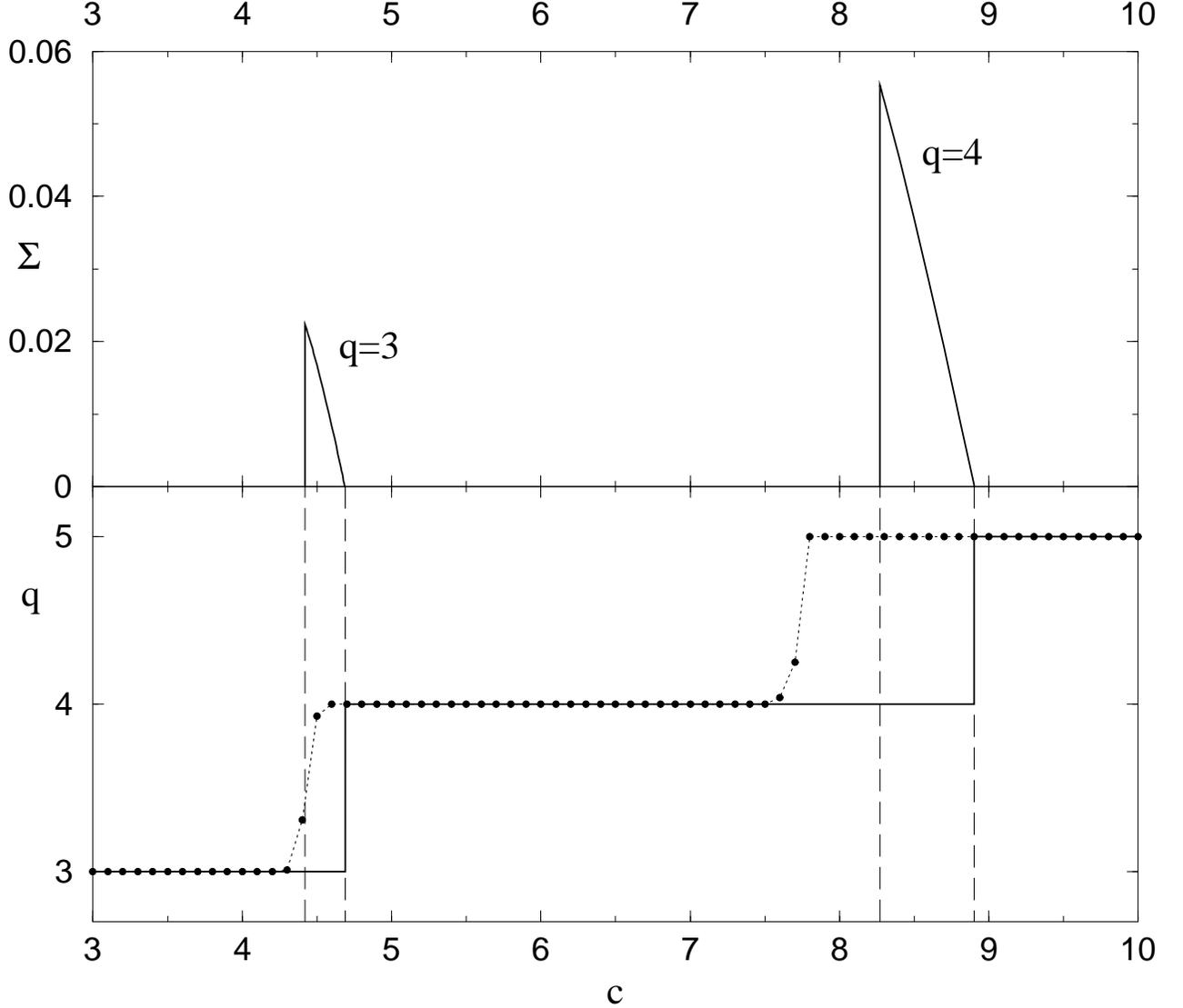}
\caption{Top: Complexity $\Sigma(c)$ vs. average connectivity for
$q=3$ and $q=4$. Non-zero complexity appears discontinuously at the
dynamical threshold $c_d$, and goes down continuously to zero at the
$q$-COL/UNCOL transition. The curves are calculated using the
population-dynamical solution for $\rho(\eta)$ with population size
${\cal M}=10^6$.\\ Bottom: The full line shows the chromatic number of
large random graphs vs. their connectivity $c$. The symbols give
results of {\it smallk} for $N=10^3$, each averaged over 100 samples.}
\label{fig:complexity}
\end{figure}

The previous analysis results for the q-coloring problem in the
existence of a dynamic transition, characterized by the sudden
appearance of an exponential number of clusters that disconnect the
solutions of the problem. This is represented in figure
\ref{fig:complexity} for $q=3$ and $4$, where the complexity is
plotted as a function of the graph connectivity.  Note, that at a
certain value average connectivity $c=c_d$ the complexity abruptly
jumps from zero to a positive value. Then it decreases with growing
$c$ and disappears at $c_q$ where the number of solutions become
zero. It is not possible any more to find a zero-energy ground state
for the system, i.e. the graph becomes uncolorable with $q$ colors,
and its chromatic number grows by one.

In the following table, we present the results for $q=3,4,$ and 5, for
the dynamical transition we show the corresponding values of $c_d$, of
the entropy $s(c_d)=\ln q + c_d \ln (1-1/q) /2$ \cite{entro_remark}
and the complexity $\Sigma(c_d)$.  For the $q$-COL/UNCOL transition,
the critical connectivity $c_q$ and the solution entropy are
given. Like in random 3-satisfiability \cite{MoZe} and vertex covering
\cite{WeHa}, this entropy is found to be finite at the transition
point.

\begin{center}
\
\begin{tabular}{|c||c|c|c||c|c|}\hline
\label{tab:entropie}
q & $c_d$ & $s(c_d)$ & $\Sigma(c_d)$ & $c_q$ & $s(c_q)$  \\ 
\hline \hline
3 & 4.42  & 0.203    & 0.0223  & 4.69 &  0.148  \\  
\hline
4 & 8.27  & 0.197    & 0.0553  & 8.90 &  0.106  \\  
\hline
5 & 12.67 & 0.196    & 0.0794  & 13.69 & 0.082  \\  
\hline 
\end{tabular}
\end{center}

\begin{figure}
\begin{center}
\includegraphics[angle=0,width=.94\columnwidth]{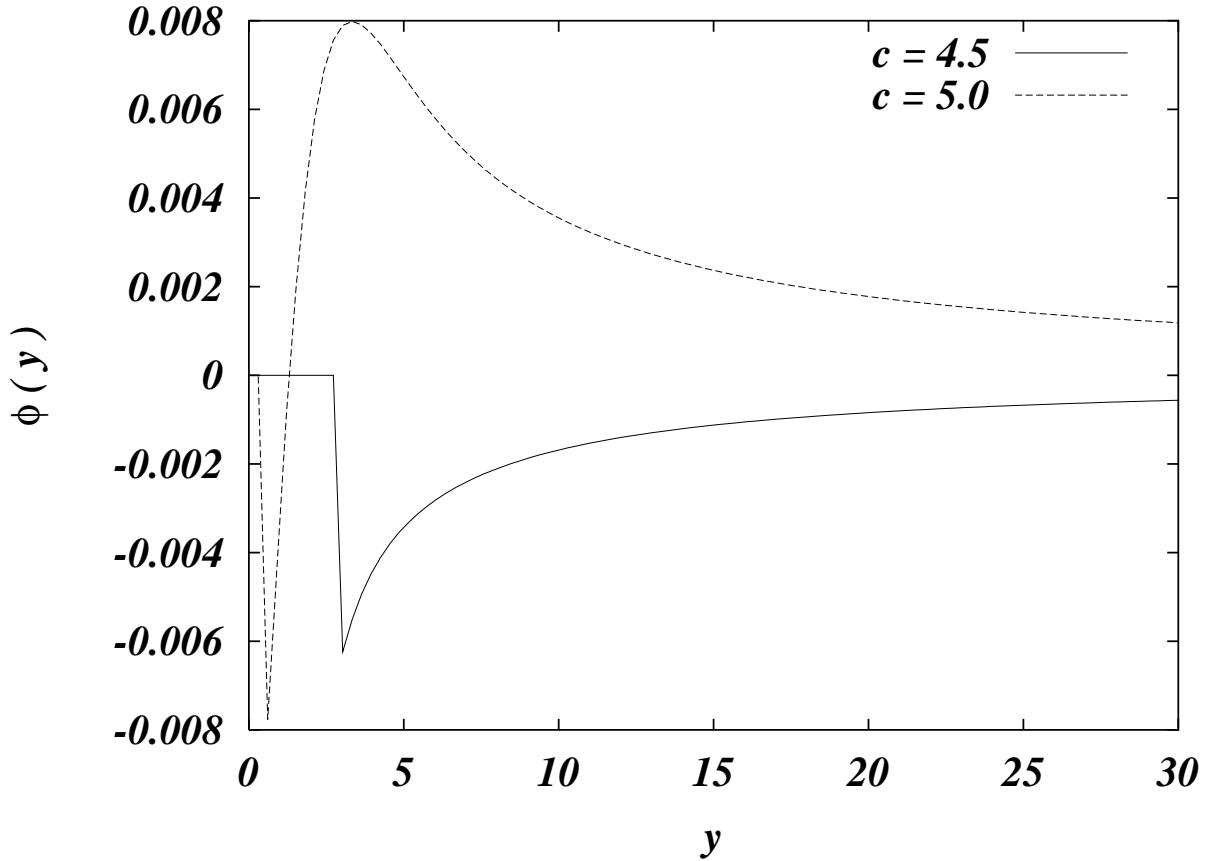}
\caption{Average thermodynamic potential $\phi(y)$ vs. $y$ in the HARDCOL
phase ($c=4.5$) and in the UNCOL phase ($c=5.0$). Note that $\phi(y)$
above the paramagnetic region ( $\phi=0$ ) is a monotonously
increasing function of $y$ in the first case, while it displays a
maximum at finite $y$ in the second one.}
\label{fig_phidiy}
\end{center}
\end{figure}

\begin{figure}
\begin{center}
\includegraphics[angle=0,width=.94\columnwidth]{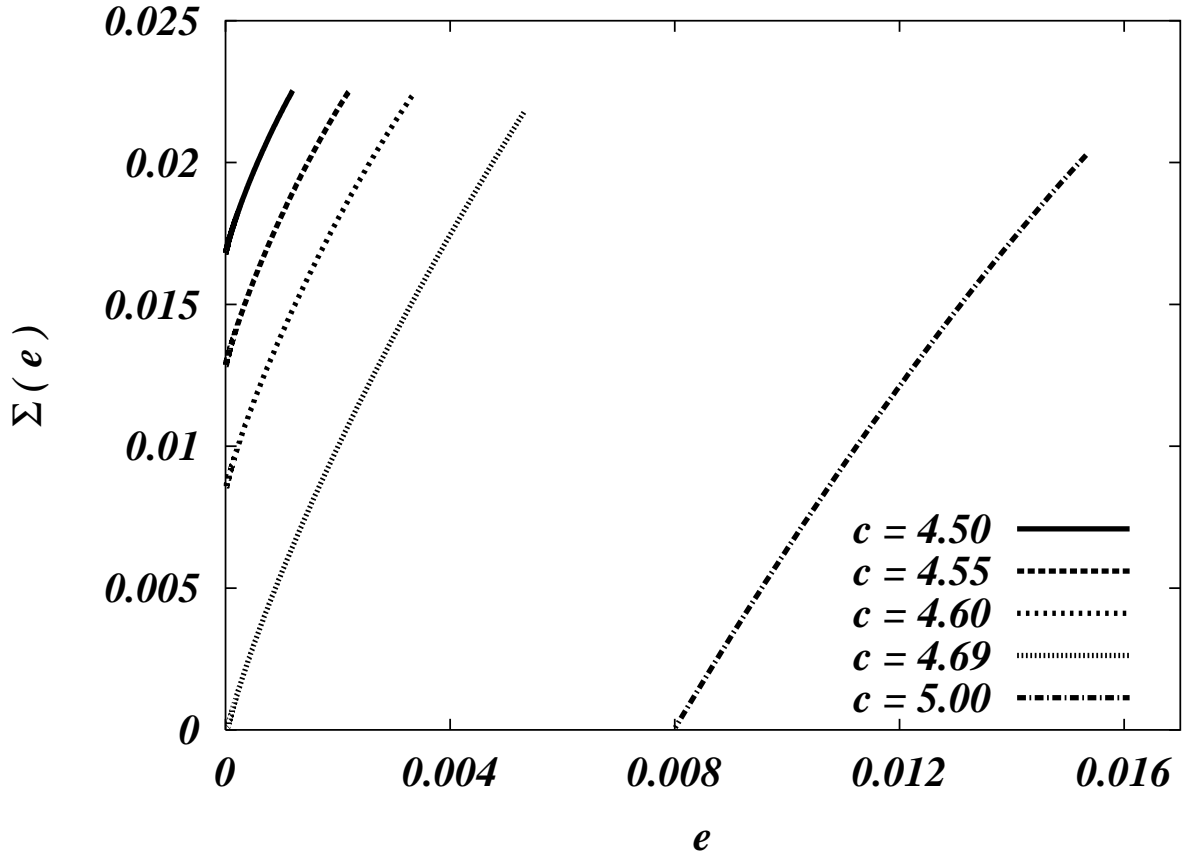}
\caption{Average complexity $\Sigma$ as a function of the energy
  density $e$ for various average connectivities $c$. In this figure
  we only display the {\em physical} branches (see text).}
\label{fig_Sigma_di_E}
\end{center}
\end{figure}

In Fig.~(\ref{fig_Sigma_di_E}) we display the average complexity
$\Sigma$ as a function of the energy density $e$ in the 1-RSB
approximation. Recently Montanari and Ricci showed in \cite{MoRi1}
that in the $p$-spin spherical spin glass the 1-RSB scheme is correct
only up to a certain critical energy density $e_G$, above which this
solution become unstable and a FRSB calculation is required. It is
possible that such a phenomenon might happen also in this case.
The dynamical transition is not only characterized by a sudden
clustering of ground states, at the same point an exponential number
of meta-stable states of positive energy appears \cite{MeZe}. Such
states (besides algorithm-dependent entropic barriers which may exist
even below $c_d$) are expected to act as traps for local search
algorithms causing an exponential slowing down of the search process.
Well known examples of search processes that are overwhelmed by the
presence of excited states are simulated annealing or greedy
algorithms based on local information.

To test this prediction, we have applied several of the best available
solvers for Coloring and SAT problems available in the net
\cite{satlib,culberson_HP}. After some preliminary simulations we
observed that the best results could be obtained with the {\em smallk}
program~\cite{culberson_HP} and concentrated our efforts on it. The
simulation results, as shown in the lower half of
Fig. \ref{fig:complexity}, were obtained in the following way: First a
random graph ($N=10^3$) was generated and we tried to color it with a
small number of colors (here $q=3$). If, after some cutoff time (we
probed with 10 seconds, 1 minute and 2 minutes without substantial
changes), the graph was not colored, we stop and tried to color it
with larger $q$.  For each connectivity we averaged over 100
samples. As it can be clearly seen, the algorithm fails with $q$
colors slightly below the dynamical transition, confirming our
expectations. In Sec. \ref{sec:SIN} we explain how the cavity approach
helps to design an algorithm being able to deal also with this problem.

\subsection{The large-$q$ asymptotic}

From Eqs. (\ref{eq_sc1rsb}) and (\ref{eq_self_cons_q}) one can
easily deduce the large-$q$ asymptotics of $\rho(\eta)$. For average
connectivities $c>>q$ (the threshold $c_q$ is expected to scale like
${\cal O}(q \ln q)$), $f_k$ is dominated by the $l=0$-contributions in
the numerator and in the denominator, leading to
$\rho(\eta)=\delta(\eta-1/q)$ in leading order. Plugging this result
into the Eq. (\ref{eq:sigma}) one can easily calculate the COL/UNCOL
threshold $c_q$ by setting the complexity to zero. Taking care only of
the dominant contribution we find
\begin{equation}
  \label{eq:cq_asympt}
  c_q = 2 q \ln q + o(q\ln q)\ .
\end{equation}
This result coincides with the exact asymptotics found by Luczak
\cite{Lu}. Note, however, that the same dominant term can also be
obtained from the vanishing of the replica-symmetric (paramagnetic)
entropy $s(c)$ which is expected to be exact up to the COL/UNCOL
transition. This means that, for $q\to\infty$, the threshold entropy
goes to zero. This behavior could already be conjectured from the
above table where the threshold entropies are given for small $q$.  
The derivation of the sub-dominant terms in Eq. (\ref{eq:cq_asympt})
requires a much more detailed analysis and goes beyond the scope of this
paper. It will be presented in a future publication together with
analogous results for $K$-SAT \cite{BrMeWeZe}.

\section{Working with single graph instances: Survey propagation}
\label{sec:SIN}

Up to now we have solved analytically the coloring problem averaged
over the set of Erd\"os-Renyi graphs at given average connectivity. In
this way we derived the $q$-dependent threshold connectivities of
$c_q$ at which the graph becomes almost surely uncolorable with $q$
colors, i.e. the location of the COL/UNCOL transition. We have also
demonstrated the existence of another threshold value $c_d$ above 
which a clustering phenomenon takes place in the space of solutions.

However, one of the relevant consequences of this cavity approach is
that it can be naturally implemented to study single case instances,
i.e. specific non-random graphs which have to have a locally tree-like
structure to fulfill the conditions of the cavity approach. In the
average-case analysis at each step of the iteration, we selected {\it
  randomly} $k$ sites from the $\cal{M}$ possible ones to be used in
Eq.~(\ref{eq_self_cons}), and we substituted another randomly chosen
entry $\eta_0$ from the ${\cal M}$ possible entries. From here on,
we will assume that the iteration procedure used above is also valid
for single instances -- with one significant change: For the
generation of survey for one vertex (or edge) we have to use its
actual neighbors, the connections between sites are fixed once for
ever by the specific graph under consideration.

\subsection{The survey propagation algorithm}

This algorithm works in a way similar to the sum product algorithm
\cite{KsFrLo}. In the latter, to each vertex arrive $u$-messages
from $k-1$ neighbors, then this messages are transformed (become
$h$-fields) and sent as a new message through the link to the descendant
$k$ neighbors. So, at each time step, in the links of the graphs you
will have messages traveling, like in a communication network. The
survey propagation algorithm (SP), works with the same principle. The
basic difference is that now the messages are replaced by $u-surveys$
of the messages (i.e. by probability distributions of messages). SP is
defined for one given value of the reweighting parameter $y$ that must
be optimized to minimize the ``free energy'' of the system. To each
edge $\{i,j\}$ of the graph we associate two u-surveys
$Q_{i\rightarrow j}(\vec{u})$ and $Q_{j\rightarrow i}(\vec{u})$ of
messages traveling in the two possible directions. The algorithm
self-consistently determines these surveys by a message passing
procedure to be described below, and finds consequently all the
thermodynamic properties of the model defined on the specific graph.
Let us now describe how SP works in practice for the 3-coloring
problem:

\begin{enumerate}
\item Select a graph $G=(V,E)$.
\item All the $Q_{i\rightarrow j}(\vec{u})$ with $\{i,j\}\in E$ are
  randomly initialized.
\item We sequentially consider all sites $i$ and randomly update the
links $\{i,j\}$ to all neighbors $j$ in the following way:
\begin{enumerate}
\item For each neighbor $j$ of $i$ we calculate:
\begin{equation}
P_{i|j}(\vec h)= C_{i|j} \int \Big[ \prod_{k \in V(i) \setminus j} d^q 
\vec u_k Q_{k\rightarrow i}(\vec u_k)\Big] \delta
\left(\vec h-\sum_{k\in V(i)\setminus j} \vec u _k\right) 
\exp \left\{ y\ \omega\left(\sum_{k\in V(i)\setminus j} \vec u_k\right)
\right\}
\label{eq:P_i-j}
\end{equation}
\noindent 
where with the symbol $V(i)$ denotes all neighbors of $i$. The
prefactor $C_{i|j}$ is chosen such that $P_{i|j}$ is properly
normalized to one.

\item From $P_{i|j}(\vec h)$ we derive the new u-surveys of all edges
  $\{i,j\}$:
\begin{equation}
Q_{i\rightarrow j}(\vec{u})=\int d^q \vec h\ P_{i|j}(\vec h)\delta
\left(\vec u- \hat u  ( \vec h ) \right)
\label{eq:P_j_i}
\end{equation}
\end{enumerate}
\item The iteration step 3. is repeated until convergence is reached.
\end{enumerate}

It was already shown in \cite{MeZe} that the free energy of the system
may be written as:
\begin{equation}
\phi(y)=\frac{1}{N} \left[ \sum_{\{i,j\}\in E}
\phi_{i,j}^{link}(y) -\sum_{i}(n_i-1)\phi_{i}^{node}(y) \right]
\label{eq:free_energy}
\end{equation}
where $n_i$ is the connectivity of the vertex $i$, and
$\phi_{i,j}^{link}(y)$ and $ \phi_{i}^{node}(y)$ represent the
contributions of links and vertices which are given by:
\begin{equation}
  \phi_{i,j}^{link}(y) = -\frac{1}{y} \ln\left( \int d^q \vec h 
  P_{i|j}(\vec h)\ d^q\vec u Q_{j\rightarrow i}(\vec u)\ 
  \exp\left\{-y\left[ \omega(\vec h) - \omega(\vec h + \vec u )
    \right] \right\} \right)
\label{eq:link_free_energy}
\end{equation}
and 
\begin{equation}
  \phi_{i}^{node}(y) = -\frac{1}{y} \ln\left( \int \prod_{k\in V(i)}
    d^q \vec u_k Q_{k\to i}(\vec u_k) \ \exp\left\{y \omega\left(
        \sum_{i=1}^k \vec u_i \right) \right\} \right)\ .
  \label{eq:site_free_energy}
\end{equation}

\begin{figure}
\includegraphics[angle=0,width=0.95\columnwidth]{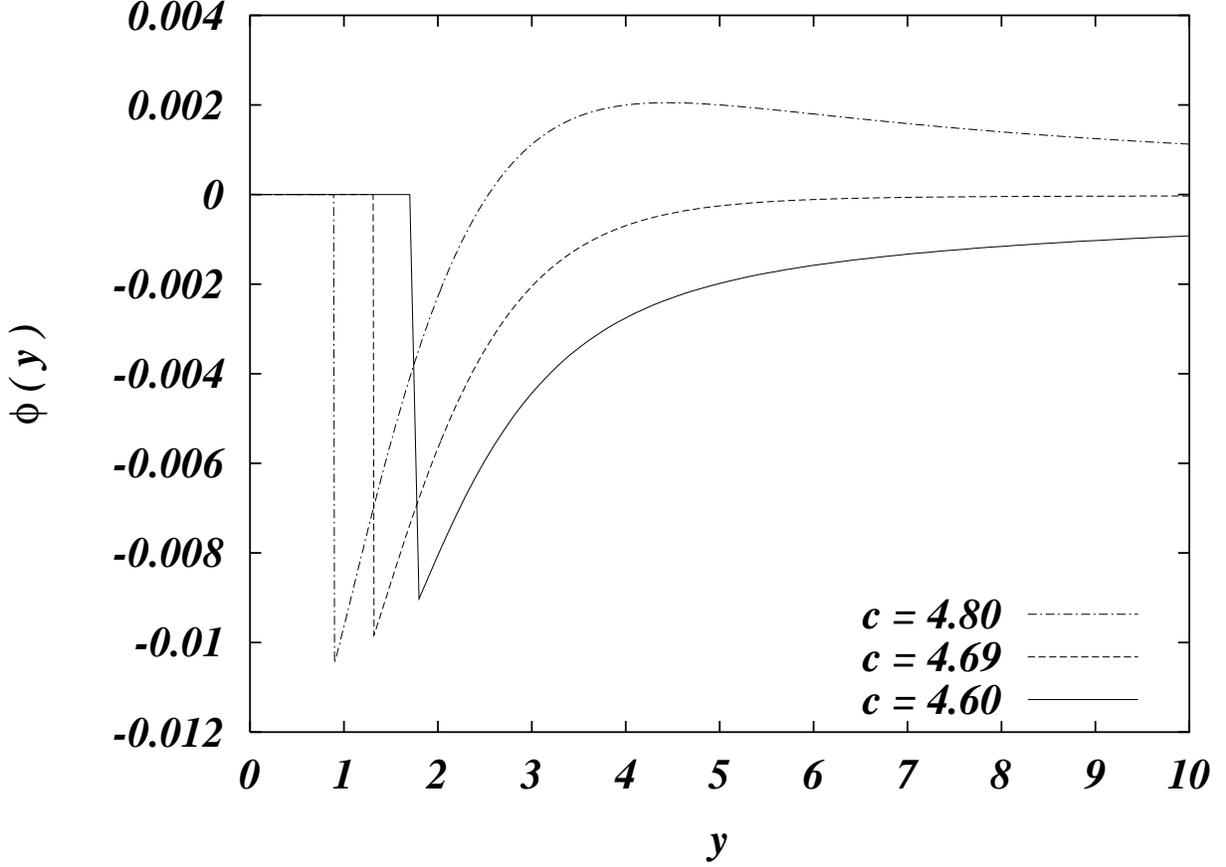}
\caption[0]{Free energies $\phi$ as a function of $y$ for three given
samples of $N=10000$ of connectivities $c=4.60,4.69,4.80$.}
\label{fig:free_energy}
\end{figure}

Repeating the above procedure for various values of $y$, Eqs.
(\ref{eq:link_free_energy}) and (\ref{eq:site_free_energy}) do not
only provide the values of $\phi(y)$, but also $\Sigma(y)= -y^2
\partial \phi(y)/\partial y$ and the energy density $e(y)=\partial(y
\Phi(y))/\partial y$ of states. The parametric plot of $\Sigma(y)$
versus $\epsilon(y)$ gives the complexity of states as a function of
their energy. For example, Fig.~(\ref{fig:free_energy}) shows the free
energy $\phi(y)$ of single graphs with $N=10000$ vertices as a function
of $y$ for three different values of the average connectivity $c$.

\begin{figure}
\includegraphics[angle=0,width=0.95\columnwidth]{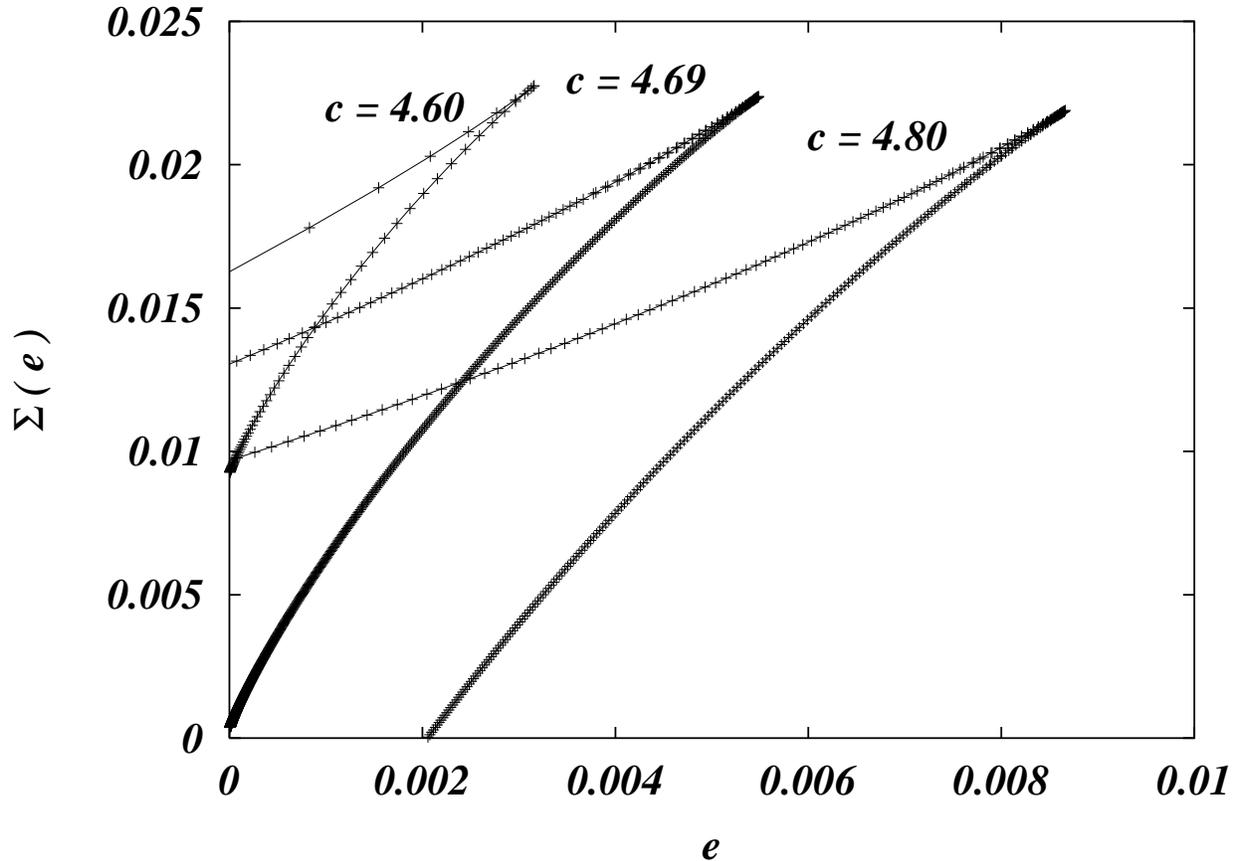}
\caption[0]{Complexity $\Sigma$ as a function of $\epsilon$ for three
given samples of random graph with average connectivities
$c=4.60,4.69,4.80$ and $N=10000$ sites. At odd with
Fig.~(\ref{fig_Sigma_di_E}) here we display both physical and
unphysical branches.}
\label{fig:complexity_e}
\end{figure}

We observe that for high enough connectivities the maximum of
$\phi(y)$ is located at finite values of $y$. While decreasing $c$,
the location of the maximum grows and approaches $y\rightarrow \infty$
at the coloring threshold. From these curves and by means of numerical
derivatives, we may also calculate the complexity and energy.
Fig.~(\ref{fig:complexity_e}) shows the two branches obtained in the
parametric plot of $\Sigma(y)$ vs. $e(y)$ for various connectivities
$c$. While the physical meaning of the upper branches is not clear
\cite{MePa2} we wanted to stress that they interpolate between the RS
solution and the maximum complexity point.

From the previous figure we may extract two characteristic values of
the energy: The first one, is associated with the minimal number $e_0
N$ of miscolored edges in the graph, i.e. it gives the {\it ground
state energy} of the instance. The value of $e_gs$ is determined as
the positive point where the lower branch of the complexity curve
intersects the energy axis, or it equals zero if $\Sigma(e=0)>0$ on
the lower branch.

The other relevant energy value is the {\it threshold energy}
$e_{th}$. It is determined by the point where the complexity reaches
its maximum. It is therefore the point where e.g. simulated annealing
gets stuck. The same remark of Sec.~({\ref{sub_res}) holds here: this
calculation should be probably improved along the line of \cite{MoRi1}
in order to take into account the FRSB instability at higher energy
density as in the case of the $p$-spin spherical model.

From the practical side this is of course not the way to determine
this values, it is much more desirable to look for the value of $y$ at
which $\phi(y)$ becomes maximal, cf. Eq.
(\ref{eq_complexity}). Fig.~(\ref{fig:energy_c}), shows a plot of
these two energies as a function of the connectivities obtained using
this single-instance algorithm.

\begin{figure}
\includegraphics[angle=0,width=0.95\columnwidth]{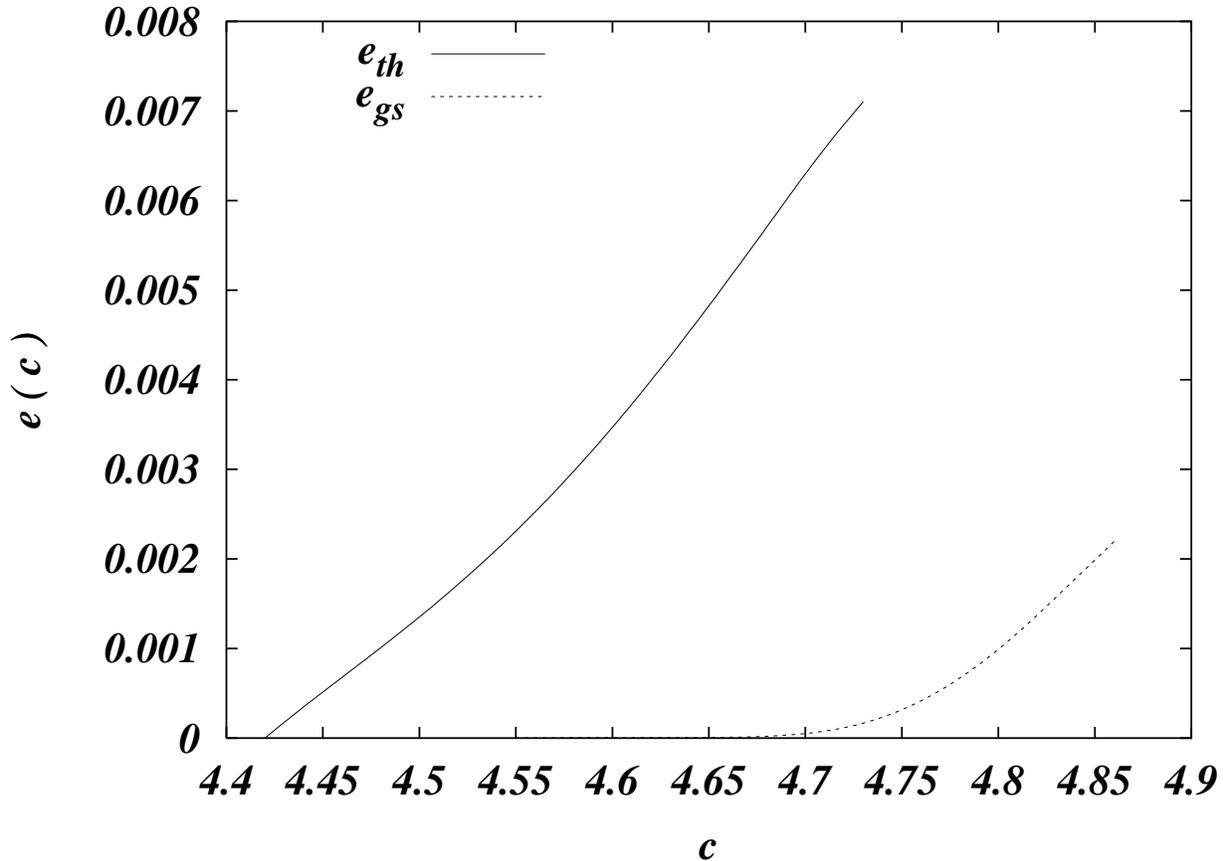}
\caption[0]{Density of miscolored links $e_{\mathrm{gs}}$ vs. average
connectivity of the graph $c$ (lower dotted curve) and threshold
energy density $e_{\mathrm{th}}$ vs. $c$ (upper continuous curve) in
the 1-RSB approximation.}
\label{fig:energy_c}
\end{figure} 

Of course, the exact meaning of the numerical values of these
quantities is an open question. In principle they were defined for
infinite systems, whereas our single-instance algorithm works for
systems of finite sizes $N$. We expect that the numerical values give
good approximations once we look to large values of $N$, where e.g.
the scales dividing distances of solutions inside one state from those
between states are well separated.  A more detailed discussion about
this may be found in \cite{MePa,MeZe}.

\section{A polynomial algorithm to color graphs}
\label{sec:ALG}

The Survey Propagation described above was very useful for the design
of an efficient algorithm to find a solution of randomly generated
3-SAT formulas \cite{MePaZe,MeZe} in the hard but satisfiable phase.
Here we will show that, with small modifications, the same idea can be
extended to the q-coloring problem.

The relevant idea in this algorithm is to fix spins which are strongly
biased toward (or away from) one color. Therefore, we have to first
determine the distributions of local magnetic fields in the system
using SP, and select those which have the strongest bias. Once these
are fixed, the problem is reduced. We can rerun SP on the reduced
instance, new spins may be biased and fixed. The procedure will be
repeated until only paramagnetic spins remain. At this point SP cannot
help any more, but surprisingly the decimated coloring problem becomes
``easy''. Using any reasonable local solver known in the literature,
we can proceed to construct a proper coloring.
				   
In the case of $q$-COL the subject is technically slightly more
complex than in K-SAT, since spins can be biased in $q$ different
directions and it is hard to decide what do we mean precisely by
biased. In addition, by fixing the color of one vertex, all its
neighbors have to have different colors, i.e. they are left with $q-1$
colors. In the reduction process the problem, initially being a pure
$q$-coloring problem, becomes a list coloring problem where each
vertex has an own list of allowed colors. In this way the permutation
symmetry of colors is broken, which requires a modification of the SP
given above to non-symmetric surveys.

In order to keep the presentation as simple as possible we concentrate
our efforts on the 3-coloring problem and hence, from now on, all the
discussion will be associated for the case $q=3$. The extension of the 
results to higher $q$ is, however, straightforward although exponential 
in $q$.

As mentioned above, the first things we should do are a generalization
of SP to non-color-symmetric situations, and to correctly define a
biased spin. Let us start first noting that equation
~(\ref{eq:dist-QQ}) may be written as 
\begin{equation}
\label{eq:dist-QQ-gen}
{\cal Q}[ Q(\vec u)] = \int d^q\vec{\eta}\ \rho(\vec{\eta})\ 
\delta\left[ Q(\vec u)- \eta^0\delta ( \vec u ) - \sum_{\tau=1}^q
\eta^{\tau} \delta (\vec u - \vec e_{\tau}) \right]
\end{equation}
where we simply avoid to consider the color symmetry of the problem,
and where we introduce $\eta^0 = (1-\sum_{\tau=1}^3 \eta^{\tau})$.
Then, following the same lines of reasoning that lead from
~(\ref{eq:dist-QQ}) to~(\ref{eq_self_cons}) we may deduce the
following update of the surveys in the limit $y\to\infty$:
\begin{equation}
\eta_{i \rightarrow j}^{r}=\frac{\prod_{k \in V(i)/j}(1-\eta_{k\to
i}^{r})-\sum_{p\neq r}\prod_{k \in V(i)/j}(\eta_{k\to
i}^{0}+\eta_{k\to i}^{p})+ \prod_{k \in V(i)/j}\eta_{k\to
i}^{0}}{\sum_{p=1,2,3}\prod_{k \in V(i)/j}(1-\eta_{k\to i }^{p})-
\sum_{p=1,2,3}\prod_{k \in V(i)/j}(\eta_{k\to i}^{0}+\eta_{k\to
i}^{p})+\prod_{k\in V(i) / j}\eta_{k\to i}^{0}} \; \; ,
\label{etaSP}
\end{equation}
for $r \in \{1,2,3\}$. The value of $\eta ^0_{i\to j }$ can be
calculated by imposing the normalization condition. Using this update
rule instead of the one proposed in the above version of SP, we
directly work with a reweighting parameter $y=\infty$ which forbids any
positive energy changes and thus characterizes proper colorings.

Having $\eta_i^\tau$, for all the sites of the graph, we have to
define the site dependent color polarizations 
\begin{equation} 
\label{eq_polarizza}
\Pi_i^r = 
\frac{\prod_{j \in V(i)}(1-\eta_{j\to i}^{r})-
\sum_{p\neq r}\prod_{j \in V(i)}(\eta_{j\to i}^{0}+
\eta_{j\to i}^{p})+ \prod_{j \in V(i)}\eta_{j\to i}^{0}}
{\sum_{p=1,2,3}\prod_{j \in V(i)}(1-\eta_{j\to i }^{p})-
\sum_{p=1,2,3}\prod_{j \in V(i)}(\eta_{j\to i}^{0}+
\eta_{j\to i}^{p})+\prod_{k\in V(i) }\eta_{j\to i}^{0}} \; \; ,
\end{equation}
for $r=1,2,3$. This equation is analogous to Eq.~(\ref{etaSP}) but the
products are extended to all neighbors. The polarization $\Pi^r_i$ is
the probability that vertex $i$ is fixed to color $r$ in a randomly
selected cluster of solutions. Vertices which may change their color
within one cluster are characterized by $\Pi^0_i =(1-\sum_{r=1}^3
\Pi^r_i)$. Once these polarizations are known, many strategies can be
adopted for coloring the graph. We believe that the simplest and most
intuitive one is the following:

\begin{itemize}
\item[(i)] If one spin is very biased to one color, fix that spin and
  remove it from the graph. Forbid this color to all neighbors.
\item[(ii)] If the bias of one spin toward some color is very low,
  forbid that color.
\end{itemize}

Forbidding a color $c$ to a node $i$ implies rewritting Eq. ~(\ref{etaSP}) 
using only two colors for that particular node. This can be achieved simply
by takin Eq. ~(\ref{etaSP}) and ~(\ref{eq_polarizza}) but setting 
$\eta_{i\to k}^c=0$ and $\eta_{k\to i}^c=1$ for all $k\in V(i)$.

Furthermore, during the processes discussed above, it turned out, that
many vertices get surrounded by neighbors with fixed colors. In that
case, the spin can be fixed to one of their remaining allowed colors
immediately, and it is removed from the graph.

In practice, we put a cutoff for the value of the bias to be used for
the previous criteria. We use rule (i) every time a bias towards some
color is greater than $0.8$ and rule (ii) if the bias was lower than
$0.15$. There is no special reason for selecting specifically these
values, but we found numerically a fast convergence to solvable
paramagnetic problem instances. It could be useful to make a
systematic analysis for improving this choice, and also to discuss
other selection rules. However, this is not the objective of the
present work. Here we just want to demonstrate that the algorithm
works substantially better than every other local search algorithm we
know, even without any parameter optimization.

Summarizing, the discussion above, our algorithm follows the next steps
\begin{enumerate}
\item Take the original graph and run SP in its infinite-$y$ version
  defined by Eq. (\ref{etaSP}).
\item Calculate the biases of all spins according to
  (\ref{eq_polarizza}). 
\item Select spins whose bias to one color is larger than 0.8, fix
  and remove these spins from the graph. Forbid the color to all
  neighbors. 
\item Select spins whose bias to one color is lower than 0.15 and
  forbid that color to these spins.
\item Take all spins where just one color is allowed, fix these spins,
  and remove then from the graph. Forbid the fixed color to all
  neighbors. 
\item If the the graph is not completely paramagnetic: rerun SP and 
  go to 2.
\item Run any smart program that solves the coloring sub-problem.
\end{enumerate}

Actually, we did not find any free program in the web which was able
to easily handle large graphs for the coloring problem. The best we
could find was the {\it smallk}-program by Culberson
\cite{culberson_HP}, but even in the easy region it exploded in memory
for graphs with sizes larger than $N=2000$. So step 7 above was
changed into:

\begin{enumerate}
\item[7.1] Transform the resulting graph into a satisfiability problem.
\item[7.2] Run walk-SAT \cite{satlib} on this satisfiability problem.
\end{enumerate}

An interesting point about the algorithm described above is the fact
that we can fix a certain percentage of spins in every algorithmic
step, without rerunning SP every time. This drastically reduces the
computational time. How many spins we may fix, depends in a
non-trivial way on the system size and on the distance from the
COL/UNCOL transition.

Figure~\ref{fig:decimation} shows the success rate of our algorithm
in 3-coloring random graphs in the hard region $c\in[4.42,4.69]$.
From left to right the sample sizes increase: $N=4*10^3$, $8*10^3$,
$16*10^3$, $32*10^3$ and $64*10^3$. In all the cases we fixed the 0.5
percentage of the spins in every iteration step. Note that keeping
this value fix we find a clear improvement of the algorithm for sizes
going from $N=4*10^3$, $8*10^3$ to $N=16*10^3$ the performance is
roughly the same for larger lattices suggesting that we should reduce
the fraction of spins to fix. However, note that even within these
conditions the algorithm works quite well in the hard region of the
system.
\begin{figure}
\includegraphics[angle=0,width=14cm]{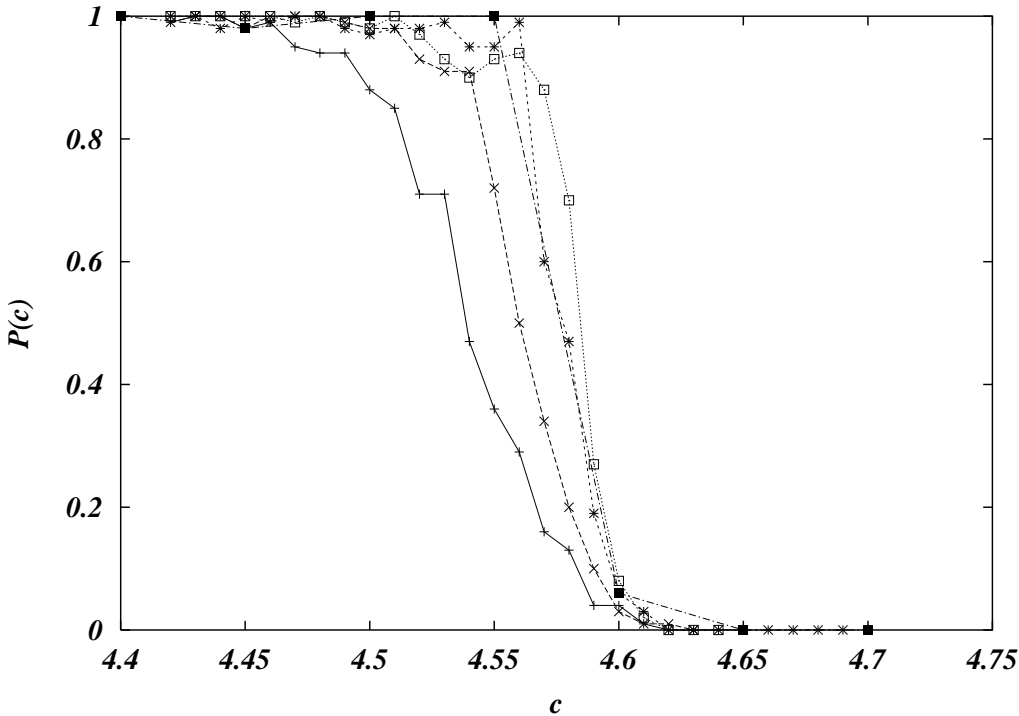}
\caption[0]{Probability of coloring a graph using our algorithm for
different lattice sizes. From left to right $N=4*10^3$, $8*10^3$
$16*10^3$, $32*10^3$ and $64*10^3$.}
\label{fig:decimation}
\end{figure} 
Note, that strong finite size effects are present, in fact the
algorithm doesn't behave very well for small lattice sizes. Two
reasons may explain this: First there are short short loops that
disappear in the thermodynamic limit, second there is some shift in
the location of the COL/UNCOL transition towards higher connectivities
for larger graphs. This point should be clarified in a fortcoming
work.

Another relevant feature of the curve is the following: The closer our
graph is to the critical point, the smaller is also the percentage of
spins we may fix in one algorithmic step. However, extrapolating the
results, the worst we can find is to fix only one single spin at the
time. This would change the complexity of our algorithm from $N \ln N$
(resulting from sorting spins with respect to their biases) to $N^2$, 
i.e. the algorithm remains polynomial.

\section{Conclusions and outlook}
\label{sec:CONC}

In this work we presented a detailed derivation of the one-step
replica-symmetry broken solution of the coloring problem on random
graphs. The problem consists in finding a coloring of all vertices of
the graph such that no two adjacent vertices carry equal colors.  From
the average case point of view, the one-step RSB approach allowed to
determine the $q$-COL/UNCOL transition $c_q$ for arbitrary color
numbers $q$. This means that large random graphs of average
connectivity below $c_q$ have proper $q$-colorings with high
probability (approaching one for $N\to\infty$), whereas graphs with
higher connectivity require more colors for a proper coloring.
Moreover, we find the existence of a clustering transition in the
colorable region. This transition is characterize by the appearance of
an exponential number of states separated by large energetic barriers.
The clustering transition is accompanied by the sudden appearance of
an exponential number of metastable states that - intuitively - cause
local algorithm to get stuck. 

We also extended our results to the study of single case instances,
i.e. specific realizations of random graphs, showing that the previous
analysis remains valid. With this understanding we also implemented a
new algorithm, based on the idea of a survey propagation that enabled
us to solve the coloring problem in the hard clustering region in
polynomial time. We present results for sizes as large as $N\simeq
10^5$ vertices, which is far beyond the performance of other
algorithms on random graphs.

There are many interesting direction to extend this work. A first one
concerns the survey-propagation algorithm. We were able to report
quite encouraging results for the SP inspired graph reduction
procedure if applied to the clustered, i.e. hard but colorable phase
on random graphs. These graphs are characterized by a locally
tree-like structure, loop are of length ${\cal O}(\ln N)$. 
This structure allowed us to use the statistical independence of
surveys restricting a randomly selected vertex inside each pure
state. This assumption fails, however, 
if the graph has some non-trivial local structure as given by small
loops, small highly connected subgraphs etc. Before being of real
practical value, SP should be extended to such situations, following
e.g. the lines used by Yedidia et al. in \cite{YeFrWe} in their
generalization of belief propagation to locally non-treelike graphs.

A second possible extension of our work concerns the interpretation of
colorings as ground-states of a Potts-antiferromagnet, which is a
model known to show glassy behavior at low temperatures (the so-called
Potts-glass), see e.g. \cite{BOOKS_SG}. In the present work we have
directly worked at zero temperature, but the extension to non-zero
temperature is straight-forward. In this context it is interesting to
see that for $q=3$ a continuous full replica-symmetry breaking
transition appears at the level of fields of ${\cal O}(T)$ - before
the one-step solution appears for fields of ${\cal O}(1)$. So we
expect that the one-step RSB transition in this model exists in a
strict sense only at zero temperature, in temperature it is only a
(sharp) cross-over to glass-like behavior. This phenomenon disappears
for larger $q$, but it is interesting in how far it can influence the
usual glassy phenomenology known from fully-connected spin-glass
models. Let us also point out that interestingly enough a similar
scenario holds also in the random K-SAT \cite{MerMezZec} case.  Using
in addition the approach suitable for single graph instances, one can,
e.g., study inhomogeneities arising in the glassy phase \cite{MoRi2}
and thus go beyond the usual paradigm of disorder averaged results for
randomly disordered models.

\begin{acknowledgments}
We are grateful to A. Montanari, J. Culberson, B. Hayes, and
F. Ricci-Tersenghi for their interest and many helpfull discussions.
\end{acknowledgments}


\begin{thebibliography}{99}
\bibitem{entro_remark} This formula is correct only in the case ${\cal
O}(T)$ fields distributions is paramagnetic (see also
Sec.~\ref{sec:CONC}). In the interval $c_d<c<c_q$ this is always true
for $q>3$.  Values of $s(c_d)$ and $s(c_c)$ displayed in the table of
Sec.~\ref{tab:entropie} for $q=3$ are therefore to be considered as a
lower bound to the true value.
\end{thebibliography}
\end{document}